\documentclass[twocolumn]{aastex63}

\turnoffediting

\received{June 20, 2020}
\revised{November 27, 2020}
\accepted{December 9, 2020}
\published{January 28, 2021}

\shorttitle{Significant Differences in \ion{H}{1} and Metals around the Lens Galaxy Towards SBS 0909+532}
\shortauthors{Cashman et al.}

\begin{document}

\title{Significant \ion{H}{1} and Metal Differences around the $z$ = 0.83 Lens Galaxy Towards the Doubly Lensed Quasar SBS 0909+532}
\vspace{10mm}

\correspondingauthor{Frances Cashman}
\email{frcashman@stsci.edu}

\author[0000-0003-4237-3553]{Frances H. Cashman}
\affil{Department of Physics \& Astronomy, University of South Carolina, Columbia, SC 29208, USA}
\affil{Space Telescope Science Institute, 3700 San Martin Drive, Baltimore, MD 21218, USA}

\author[0000-0002-2587-2847]{Varsha P. Kulkarni}
\affil{Department of Physics \& Astronomy, University of South Carolina, Columbia, SC 29208, USA}

\author[0000-0003-0389-0902]{Sebastian Lopez}
\affil{Departamento de Astronom\'{i}a, Universidad de Chile, Casilla 36-D, Santiago, Chile}
\vspace{10mm}

\begin{abstract}
We report a large difference in neutral hydrogen (\ion{H}{1}) and metal column densities between the two sight lines probing opposite sides of the lensing galaxy at $z_\mathrm{lens}$ = 0.83 toward the doubly lensed quasar SBS 0909+532. Using archival \textit{HST}-STIS and \textit{Keck} HIRES spectra of the lensed quasar images, we measure log $N_\mathrm{H\;I}$ \edit1{= 18.77 $\pm$ 0.12 cm$^{-2}$} toward the brighter image ($A$) at an impact parameter of $r_A$ = 3.15 kpc and  log $N_\mathrm{H\;I}$ = 20.38 $\pm$ 0.20 cm$^{-2}$ toward the fainter image ($B$) at an impact parameter of $r_B$ = 5.74 kpc. This difference by a factor of \edit1{$\sim$41} is the highest difference between sight lines for a lens galaxy in which \ion{H}{1} has been measured, suggesting patchiness and/or anisotropy on these scales. We estimate an average Fe abundance gradient between the sight lines to be \edit1{$\geq$ +0.35} dex kpc$^{-1}$. The $N_\mathrm{Fe\;II}$/$N_\mathrm{Mg\;II}$ ratios for the individual components detected in the \textit{Keck} HIRES spectra have supersolar values for \edit1{all components in sight line $A$ and for 11 out of 18 components in sight line $B$}, suggesting that Type Ia supernovae \edit1{may have contributed} to the chemical enrichment of the galaxy's environment. Additionally, these observations provide complementary information to detections of cold gas in early-type galaxies and the tension between these and some models of cloud survival.
\end{abstract}


\section{Introduction} 
\label{sec:intro}
Traditional quasar absorption line studies probe a single sight line through a galaxy; however, it is difficult to link the properties of an absorber to the galaxy host and speculate about the galaxy's properties based on a single sight line. The use of gravitationally lensed quasars (GLQs) to probe foreground galaxies improves on the single sight line method, as one has multiple sight lines to characterize the absorption regions of the galaxy. Using multiple sight lines has the advantage of studying variations in gas, dust, and structure to offer a unique transverse study of a galaxy. Locally, multiple sight lines have been successfully implemented to probe the interstellar medium (ISM) of the Milky Way (MW) and other nearby galaxies. \cite{2000ApJ...543L..43L} and \cite{2001ApJ...552L..73A} revealed turbulence-driven astronomical unit-scale variations in cold neutral gas structures traced by low column density \ion{Na}{1} absorption lines along closely spaced stellar sight lines. Similarly, closely spaced sight lines toward GLQs can distinguish  small-scale structure in the ISM of a lens galaxy. Additionally, any absorbers that exist between the lens and the background quasar are magnified by lensing, potentially revealing parsec- to kiloparsec-scale structure depending on the location of the absorber.

Absorption line studies of lenses are not as common as non-lens absorbers since most lenses lie at a redshift $z \lesssim$ 1 and determining metallicities requires measurement of the \ion{H}{1} column density, which requires UV spectroscopy (as does any absorption line system with $z <$ 1.6). Even though the opportunity to study lenses in the UV is rare due to limited access to instrumentation, it is a task worth insisting on, as analysis of GLQ images permits not only the determination of galaxy mass but also abundance gradients within the galaxy. This is noteworthy, as some lenses have shown positive or inverted gradients, i.e., gradients opposite to what is seen in the MW and other nearby galaxies, suggesting central dilution from mergers or infall from metal-poor gas ($\sim$ $-$0.01 to $-$0.09 dex kpc$^{-1}$ in the MW, \citealt{2002AJ....124.2693F}; \citealt{2011AJ....142..136L}; \citealt{2012ApJ...746..149C};  $-$0.043 dex kpc$^{-1}$ in M101, \citealt{2003ApJ...591..801K};  $-$0.027 $\pm$ 0.012 dex kpc$^{-1}$ in M33, \citealt{2008ApJ...675.1213R}; $-$0.041 $\pm$ 0.009 dex kpc$^{-1}$ in nearby isolated spirals, \citealt{2010ApJ...723.1255R})

Previous lens galaxy imaging surveys suggest that the majority of lens galaxies are passively evolving normal early-type galaxies (e.g., \citealt{1998ApJ...509..561K}; \citealt{2016MNRAS.458.2423Z}). \cite{1998ApJ...509..561K} describe that lens galaxies are a biased sample, typically very massive, as massive galaxies are more likely to lens background objects. This mass bias favors early-type galaxies, with late-type spirals expected to compose only $10-20\%$ of all gravitational lenses. It has also been reported that approximately a third of nearby early-type ellipticals are not gas-poor but contain large amounts of \ion{H}{1} gas, despite being quiescent (e.g., recent 21 cm surveys by \citealt{2009A&A...498..407G}; \citealt{2010MNRAS.409..500O}; \citealt{2012MNRAS.422.1835S}; \citealt{2014MNRAS.444.3408Y}). Additionally, there are QSO absorption line studies of luminous red galaxies in which \ion{Mg}{2} ${\lambda\lambda}$2796, 2803 has been detected at distances $\gtrsim$100 kpc (e.g.,  \citealt{2010ApJ...716.1263G, 2009ApJ...702...50G}; \citealt{2016MNRAS.455.1713H}), suggesting that these halos are chemically enriched. This detection of enriched cool gas within passive galaxies raises questions as to what processes and mechanisms exist within the galaxy that keep the gas from cooling further and forming stars. 

The mass bias in lensing studies can be advantageous, as studying variations within these lensing galaxies along sight lines to the multiple images can contribute significantly to what we know about the ISM of passively evolving elliptical galaxies. In a QSO absorption line study along multiple sight lines to three lensing galaxies, \cite{2016MNRAS.458.2423Z} and \cite{2017ApJ...846L..29Z} reported that while the gas content varied amongst the lenses and within sight lines of the same lenses, supersolar [Fe/Mg] relative abundance patterns were observed in all sight lines that also had detections of cool gas. The high [Fe/Mg] ratios suggest a significant contribution from Type Ia supernovae (SNe Ia) to the chemical enrichment history of the inner ISM of these lenses. These observations support current theories that the presence of mature stellar populations could prevent further star formation from occurring in the reservoirs of chemically enriched cool gas due to a combination of injected energy from SNe Ia and winds from asymptotic giant branch stars (e.g., \citealt{2015ApJ...803...77C}).

We used public archive \textit{HST}-STIS UV absorption spectra and \textit{Keck} HIRES optical spectra to study the unexplored difference in the chemical composition and kinematic structure of the lens galaxy at $z = 0.83$ toward the double GLQ SBS 0909+532. We report significant \ion{H}{1} and metal column density differences at projected distances from the lens galaxy's center of $r_A$ = 3.15 and $r_B$ = 5.74 kpc on opposite sides of the lens in the inner ISM. We see much heavier \ion{H}{1} and metal absorption in image \textit{B} than in image \textit{A}. We discuss the observed $N_\mathrm{Fe\;II}$/$N_\mathrm{Mg\;II}$ relative abundance pattern in SBS 0909+532 in comparison to the lenses Q1017$-$2046, Q1355$-$2257, Q0047$-$1756, and HE 0512$-$3329 from the literature. Additionally, we report coincident \ion{Mg}{2} absorption at a redshift of $z_\mathrm{abs}$ = 0.611 along both lines of sight in \S\ref{subsec:abund06}.

This paper is organized as follows. In section \ref{sec:obs}, we describe the archival observations and how the data were reduced. In section \ref{sec:colden}, we describe how the column density measurements and metallicities were determined for images \textit{A} and \textit{B}. In section \ref{sec:results}, we show the results of our measurements. In section \ref{sec:disc} we compare our results to similar studies from the literature. We adopt the cosmology $H_{0}$ = 70 km s$^{-1}$ Mpc$^{-1}$, ${\Omega}_{M}$ = 0.3, and ${\Omega}_{\Lambda}$ = 0.7 throughout this paper. 

\section{Observations and Data Reduction} 
\label{sec:obs}
\subsection{SBS 0909+532 Background}\label{subsec:background}
SBS 0909+532 is a well-studied doubly imaged gravitationally lensed system identified by \cite{1997ApJ...479..678K}, who initially observed two images at $z_\mathrm{QSO}$ = 1.377 separated by 1\farcs1. \cite{1997ApJ...491L...7O} were able to confirm the redshift of the source quasar, as well as the redshift of the reported \ion{Mg}{2} absorption seen at $z = 0.83$. The lens redshift was later spectroscopically confirmed to be $z$ = 0.8302 $\pm$ 0.0001 by \cite{2000AJ....119..451L}, who also remeasured the quasar's redshift at $z = 1.3764\pm0.0003$. \cite{2000ApJ...536..584L} conducted high angular resolution HST imaging and concluded that the lensing galaxy is, morphologically, a normal early-type galaxy with an extended dark matter halo with a large effective radius ($r_\mathrm{eff}$ = 1$\farcs58\pm0\farcs$90) and low surface brightness. \cite{2000AJ....119..451L} estimated the mass inside the Einstein ring to be 1.42 $\times$ 10$^{11}$ $M_{\sun}\;h^{-1}$. The core of the lens is observable in $I$-band imaging as a significant residual after the quasar images are subtracted, but it does not appear in the $V$ band. The local environment of the lensing galaxy consists of three nearby galaxies within $\sim$200 kpc $h^{-1}$, two of which are within $\sim$100 kpc $h^{-1}$. 

The difference in magnitudes of the $A$ and $B$ images inferred from photometry, as well as low-resolution spectroscopy, of SBS 0909+532 shows an extinction curve produced by dust in the lens galaxy (\citealt{2002ApJ...574..719M}), the first optical extinction curve measurement at cosmological distances that matches the quality seen locally. We note that differential extinction has also been detected in another galaxy at $z$ = 0.93 (\citealt{2003A&A...405..445W}). Both \cite{2002ApJ...574..719M} and \cite{2005ApJ...619..749M} suggest that there may be a unique link between the activity of a galaxy and the strength of the 2175 {\AA} feature since passive, normal early-type galaxies are subjected to fewer shocks and processing by radiation, unlike the environments of starburst galaxies and active galactic nuclei. 

What has been missing from the portrait of SBS 0909+532 is a description of the variation in metallicity and kinematic structure between the two sight line images. As many lens galaxies lie at $z < 1$, the hydrogen Lyman series lines fall in the UV portion of the electromagnetic spectrum and are therefore only accessible to study with UV spectrographs on board space-based telescopes. Therefore, measurements of \ion{H}{1} exist for only a handful of lenses, and this, in turn, means that metallicity measurements of lens galaxies are also scarce. In this paper, we use archival \textit{HST} and \textit{Keck} spectroscopy measurements to provide a deeper probe into the environment of early-type lens galaxies by providing multiple metallicity measurements at different impact parameters through the ISM of the lens galaxy. Only four other measurements of \ion{H}{1} exist for lens galaxies (Q1017$-$2046 \textit{AB}, Q1355$-$2257 \textit{AB}, \citealt{2019ApJ...886...83K}; Q0047$-$1756 \textit{AB}, \citealt{2017ApJ...846L..29Z}; Q0512$-$3329 \textit{AB}, \citealt{2005ApJ...626..767L}; all lenses with redshift $0.48 < z < 1.1$), and three of the four lenses (Q1017$-$2046 \textit{AB}, Q1355$-$2257 \textit{AB}, Q0512$-$3329 \textit{AB}) show a positive average metallicity gradient (or inverse gradient).

\begin{figure*}[htb!]
\epsscale{0.7}
\plottwo{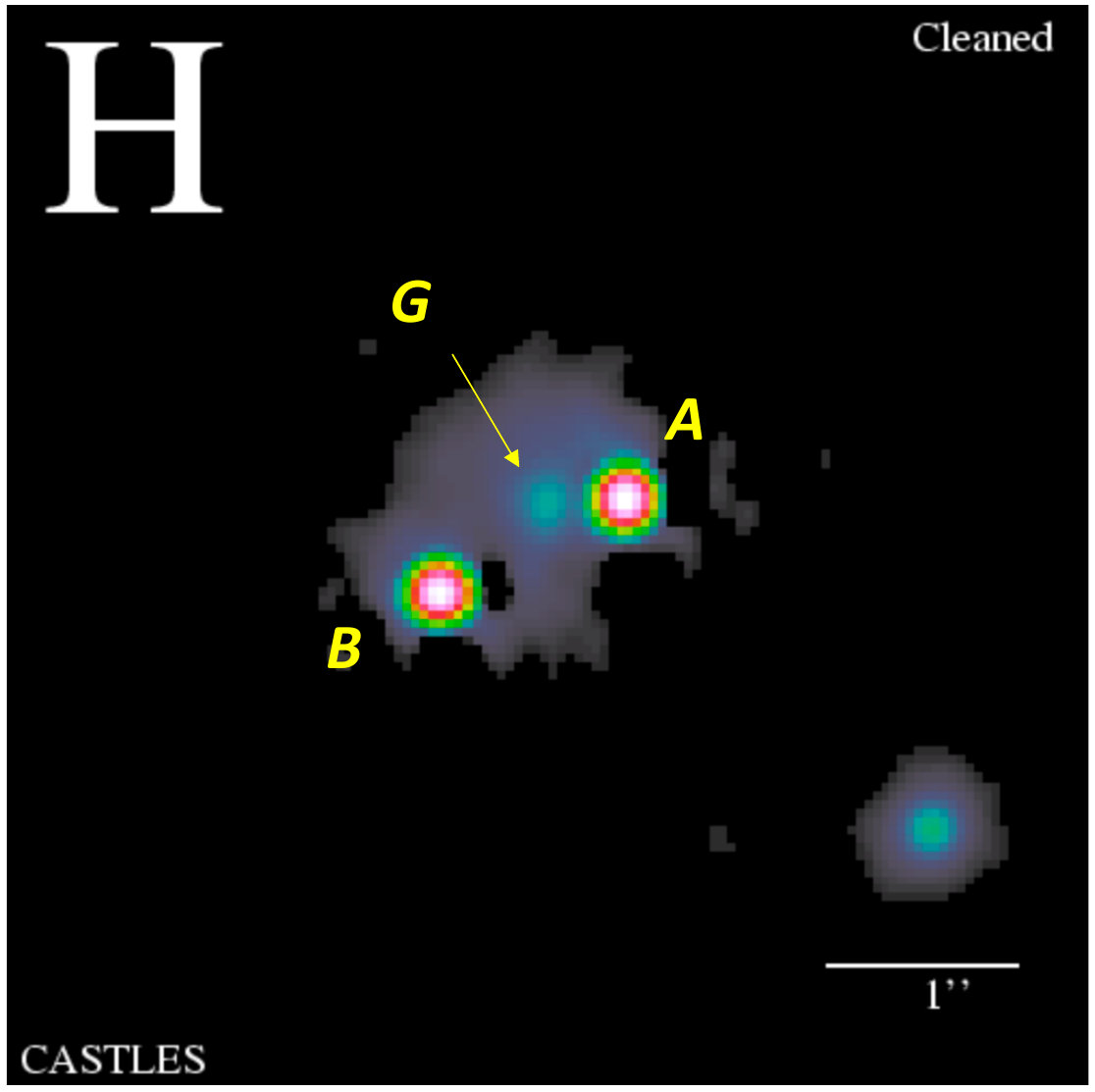}{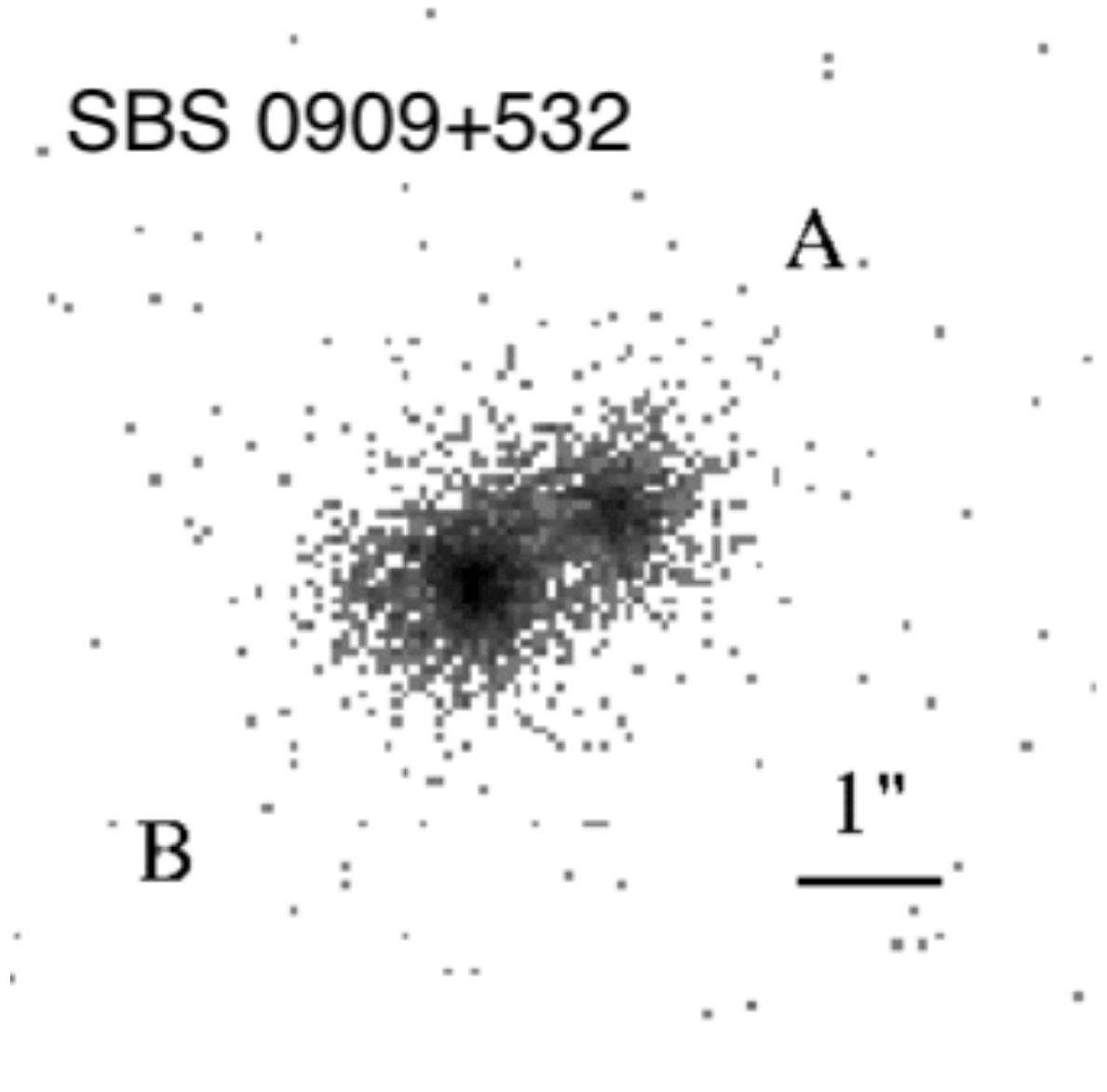}
\caption{\textit{Left}: HST image of SBS 0909+532 \textit{AB} in the $H$ band from the CASTLeS Survey (\citealt{1999AIPC..470..163K}). The lens galaxy is clearly visible in between the lensed images. \cite{2000ApJ...536..584L} report an absorption-corrected optical flux ratio in the $H$ band of $A/B$ = 1.12. \textit{Right}: \textit{Chandra} Observations of SBS 0909+532 \textit{AB} from \cite{2009ApJ...692..677D}.
\label{fig:glqAB}}
\end{figure*}

A summary of the observational and spectroscopic parameters for the QSO SBS 0909+532 and the absorption line systems detected along the line of sight are shown in Tables \ref{tab:0909param} and \ref{tab:absorbers}. The impact parameters $r$, i.e., the projected positions of the lensed images \textit{A} and \textit{B} relative to the lens galaxy, are visually marked in Figure \ref{fig:glqAB}. Image \textit{A} is the brighter image and has a projected distance of 3.15 kpc from the lens galaxy. Image \textit{B} is fainter and is at a projected distance of 5.74 kpc from the lens galaxy.

\begin{deluxetable*}{cccccccc}[htb!]
\tabletypesize{\scriptsize}
\tablewidth{0pt} 
\tablecaption{Basic Properties of the QSO SBS 0909+532 $AB$ and the Lens Galaxy
\label{tab:0909param}}
\tablehead{
\colhead{$z_\mathrm{QSO}$} & \colhead{$z_\mathrm{lens}$} & \colhead{Mag$_A$, Mag$_B$\tablenotemark{a}} & \colhead{$\Delta\theta_{AB}$ (arcsec)\tablenotemark{b}} & \colhead{$r_A$ (kpc)\tablenotemark{c}} & \colhead{$r_{B}$ (kpc)\tablenotemark{d}} & \colhead{$l_{A,B}$ (kpc)\tablenotemark{e}} 
}
\startdata
1.3764 $\pm$ 0.0003 & 0.8302 $\pm$ 0.0001& 16.07, 16.42 ($I$) & 1.17 & 3.154 & 5.744 & 8.9 \\
\enddata
\tablenotetext{a}{$I$-band magnitude of each lensed quasar image (\citealt{2000ApJ...536..584L})} 
\tablenotetext{b}{Angular separation between lensed quasar images}
\tablenotetext{c}{Impact parameter of image \textit{A} from the lens galaxy, i.e., the projected distance between sight line \textit{A} and the galaxy center}
\tablenotetext{d}{Impact parameter of image \textit{B} from the lens galaxy, i.e., the projected distance between sight line \textit{B} and the galaxy center}
\tablenotetext{e}{Transverse separation between the GLQ sight lines at the lens redshift}
\end{deluxetable*}

\begin{deluxetable}{cccc}[htb!]
\tabletypesize{\scriptsize}
\tablewidth{0pt} 
\tablecaption{Absorption Line Systems Observed along the Lines of Sight to SBS 0909+532 \label{tab:absorbers}}
\tablehead{
\colhead{$z$} & \colhead{$W^\mathrm{2796}_{A}$ (\AA)\tablenotemark{a}} & \colhead{$W^\mathrm{2796}_{B}$ (\AA)\tablenotemark{b}} & \colhead{$l_{A,B}$ (kpc)\tablenotemark{c}} 
}
\startdata
0.611   &  0.14 $\pm$ 0.01 & 0.36 $\pm$ 0.01 & 7.9 \\
0.8302 & 0.09 $\pm$ 0.02  & 2.63 $\pm$ 0.01 & 8.9 \\
\enddata
\tablenotetext{a}{Equivalent width of the \ion{Mg}{2} $\lambda$2796 absorption line in sight line \textit{A}}
\tablenotetext{b}{Equivalent width of the \ion{Mg}{2} $\lambda$2796 absorption line in sight line \textit{B}}
\tablenotetext{c}{Transverse separation between the GLQ sight lines at the absorber redshift}
\end{deluxetable}

\subsection{Keck HIRES Observations and Data Reduction}
\label{subsec:keck}
Spatially resolved spectra of SBS 0909+532 \textit{A} and \textit{B} were downloaded from the \textit{Keck} Observatory archive\footnote{\url{https://www2.keck.hawaii.edu/koa/public/koa.php}}. These data were obtained with the HIRES spectrograph (single CCD setup) on the night of 1998 December 18 as part of program C140H (PI: W. Sargent) on the \textit{Keck I} telescope. Five 2700 s exposures were taken of image \textit{A} and 12 3000 s exposures were taken of image \textit{B}. These data were reduced using the \texttt{MAKEE}\footnote{\url{https://www.astro.caltech.edu/~tb/makee/index.html}} package developed by T. Barlow. \edit1{The 2-pixel projected slit used was $\sim$0.6$\arcsec$ wide as projected on the sky, which yielded a spectral resolving power $R$ $\sim$54,000.} The extracted one-dimensional spectra were binned in the spectral direction by two. Continuum-normalized spectra were produced using the Python program \texttt{linetools} \citep{2017zndo...1036773P} by fitting a spline function to the continuum of each quasar image. The wavelength coverage was $\sim$$4300-6300$ {\AA}. The one-dimensional spectra were then co-added for each image.  

\subsection{Hubble Space Telescope STIS Observations and Data Reduction}
\label{subsec:stis}
SBS 0909+532 was observed with \textit{HST}-STIS (Space Telescope Imaging Spectrograph) on 2003 March 7 in both the optical and the UV using the CCD detector and the G430L and G230LB gratings with STIS during Cycle 11 as part of program GO-9380 (PI: E. Mediavilla). The data were initially presented in \cite{2005ApJ...619..749M} and were used to determine an optical-FUV differential extinction curve of the dust in the lens galaxy. We refer the reader to that paper for more specific details on the observation.

The \ion{H}{1} column density is needed to measure the gas metallicity. The lens galaxy at $z$ $\sim$0.83 results in the hydrogen Ly$\alpha$ line being redshifted to $\lambda \sim$2224.7 {\AA}, which appears in the G230LB data with wavelength coverage $1667-3070$ {\AA}. The reduced two-dimensional G230LB data were examined to determine the locations of the traces of the two resolved quasar images and were further processed using the \texttt{IRAF}\footnote{\texttt{IRAF} is distributed by the National Optical Astronomy Observatory (operated by the Association of Universities for Research in Astronomy Inc.) under cooperative agreement with the National Science Foundation}\texttt{/STSDAS X1D} \citep{1993ASPC...52..173T} task to extract the one-dimensional spectra. Background subtraction, charge transfer efficiency correction, conversion to heliocentric wavelengths, and absolute flux calibration were also performed during the \texttt{X1D} task. The average wavelength dispersion was 1.37 {\AA}  pixel$^{-1}$, the average resolution was $\sim$500 km s$^{-1}$, and the S/N ratio per pixel is $\sim$30 in image \textit{A} and $\sim$16 in image \textit{B} in the regions of interest. As the flux levels between the exposures were consistent, the three exposures of \textit{A} were combined with equal weighting. During the combination of exposures, gaps in spectral coverage due to bad pixels were recovered by replacing the pixel with the average flux value from the other two frames. The same procedure was performed for image \textit{B}. The combined exposures of \textit{A} and \textit{B} were then continuum normalized using \texttt{linetools} by fitting a spline function in featureless spectral regions. 

\section{Absorption Line and Column Density Measurements}
\label{sec:colden}
Where possible, two methods were used to measure column densities: the apparent optical depth (AOD) method (\citealt{1996ARA&A..34..279S}) using the program \texttt{SPECP}\footnote{\texttt{SPECP} was developed by D. Welty and J. Lauroesch.} and Voigt profile fitting using the program \texttt{VPFIT} version 11.1 \citep{2014ascl.soft08015C}. The atomic data utilized by \texttt{VPFIT} and \texttt{SPECP} were adopted from the compilations of \cite{2017ApJS..230....8C} and \cite{2003ApJS..149..205M} (see Table \ref{tab:lenstabA} notes $a$-$d$ for atomic information on the specific transitions included in the fits).

\subsection{STIS \ion{H}{1} Measurements}
\label{stis}
\edit1{The \ion{H}{1} Ly$\alpha$ and Ly$\beta$ lines were used for estimating the column density along the sight line to the brighter image $A$. Unfortunately, all shorter wavelength Lyman series lines, although covered,} are either in regions of extreme noise or are severely blended with features at different redshifts (Ly forest lines at lower redshifts and MW ISM lines). \edit1{The only \ion{H}{1} Lyman series line that could be used for estimating the column density along the sight line to the fainter image $B$ was the \ion{H}{1} Ly$\alpha$ line.} Although the column \edit1{density in sight line $B$ is estimated without these higher order Lyman series lines (possibly within orders of magnitude), we determine a reasonably robust value for log $N_\mathrm{H\;I}$. Of course, we note} that observations of SBS 0909+532 \textit{AB} at higher resolution in the \edit1{far-UV could permit measurements of higher order Lyman series lines up to the \ion{H}{1} Lyman limit}. The \ion{H}{1} column densities for \textit{A} and \textit{B} were measured based on single component fits to the Ly$\alpha$ \edit1{and/or Ly$\beta$} lines using \texttt{VPFIT}. The methods to obtain the \ion{H}{1} column densities are described below.

\subsubsection{SBS 0909+532 \textit{A}}
\label{subsec:stisA}
\edit1{We find weak \ion{H}{1} absorption near $z_\mathrm{lens}$ = 0.83 in the spectrum of SBS 0909+532 \textit{A}. As can be seen in the left panels of Figure \ref{fig:hiAB} and in Figure \ref{fig:hiab_oplot}, the \ion{H}{1} Ly$\alpha$ feature spans a velocity range from $\sim-600$ to +600 km s$^{-1}$, and although it is broad and saturated, at this resolution ($v_\mathrm{FWHM}$ = 333 km s$^{-1}$) it does not display damping wings. Since both Ly$\alpha$ and a partially blended Ly$\beta$ were detected, they were fit together to constrain the \ion{H}{1} column density. We chose a conservative estimate for the $b$-value of the single component to be $v_{\mathrm{FWHM}}$/10 (i.e., $b$ = 33 km s$^{-1}$), based on Legendre-Gauss quadrature representing the integration over the instrumental profile. The resulting log $N_\mathrm{H\;I}$ from fitting both lines with \texttt{VPFIT} is 18.77 $\pm$ 0.12 cm$^{-2}$ and the profile shows good agreement with the data (see the left panels of Figure \ref{fig:hiAB}). Repeating the fit with a $b$-value as small as 12 km s$^{-1}$ results in log $N_\mathrm{H\;I}$ = 18.83 $\pm$ 0.09 cm$^{-2}$, which is within the margin of error of our log $N_\mathrm{H\;I}$ measurement based on the instrument profile. Although such a small $b$-value cannot be justified given the instrument profile, we include this result to show the robustness of the estimate corresponding to a conservative $b$-value based on the data resolution. We performed an additional fit for a $b$-value as large as 45 km s$^{-1}$ resulting in log $N_\mathrm{H\;}$ = 18.60 $\pm$ 0.17 cm$^{-2}$. This result is consistent with our conservative measurement at about the 1$\sigma$ level. These fits can be seen in the left panels of Figure \ref{fig:hiAB}.}

\subsubsection{SBS 0909+532 \textit{B}}
\label{subsec:stisB}
\begin{figure*}[htb!]
\centering
\epsscale{0.57}
\plotone{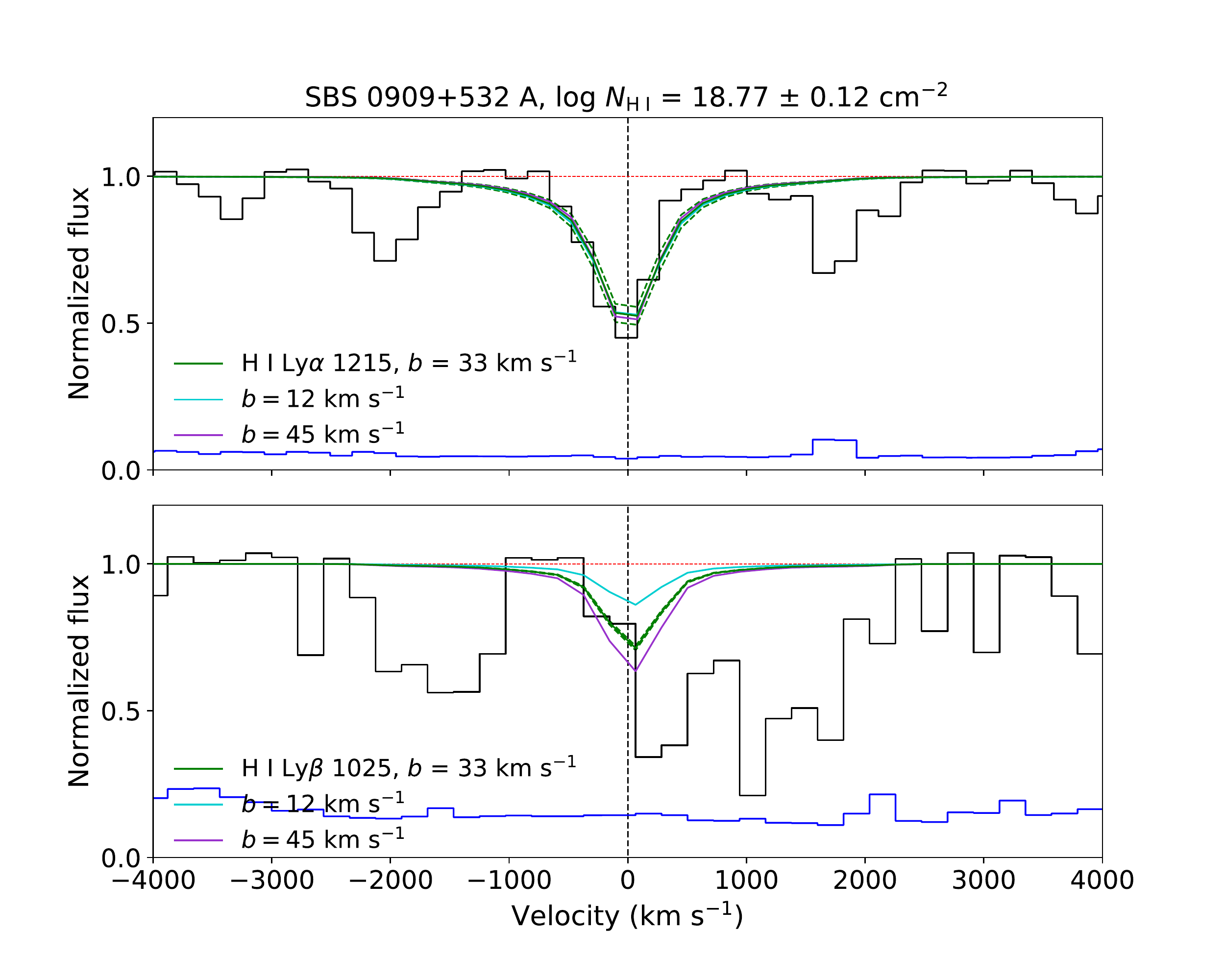}
\plotone{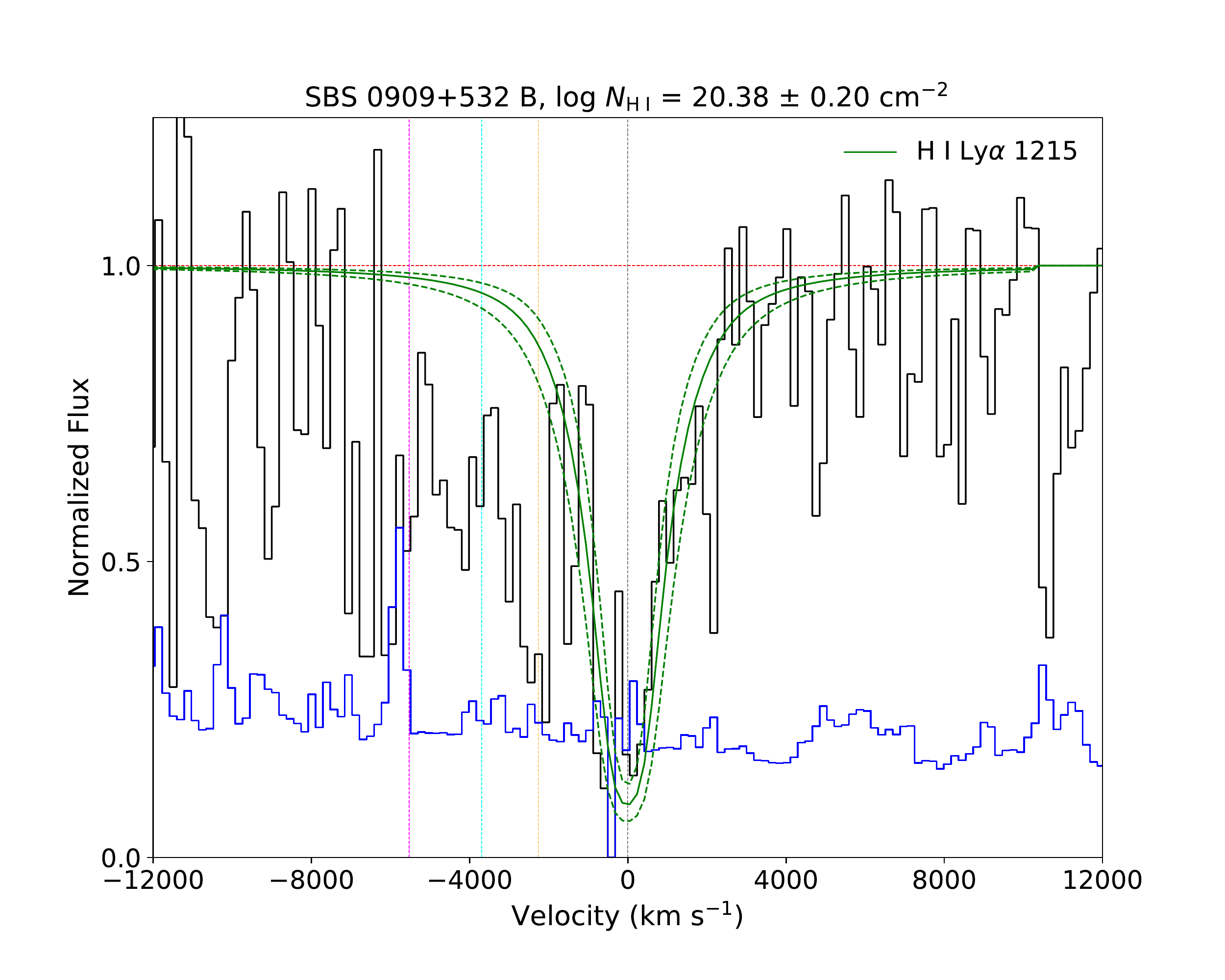}
\caption{\edit1{Plots of the detected \ion{H}{1} line(s) in the $z = 0.83$ lens galaxy in the \textit{HST} STIS spectrum of SBS 0909+532 $A$ and $B$. In each panel, the normalized data are shown in black, the dashed red line shows the continuum level, and the blue curve near the bottom shows the the 1${\sigma}$ error in the normalized flux. The solid green curve in each panel indicates the theoretical Voigt profile fit. The dashed green curve above and below the fitted profile shows the uncertainty in log $N_\mathrm{H\;I}$. The black vertical dashed line indicates the position of the \ion{H}{1} component that was used in the fit (see Tables \ref{tab:lenstabA} and \ref{tab:lenstabB}). Left panels: The Voigt profile fits for \ion{H}{1} Ly$\alpha$ and Ly$\beta$ corresponding to log $N_\mathrm{H\;I}$ = 18.77 $\pm$ 0.12 cm$^{-2}$ with $b$ = 33 km s$^{-1}$ in the \textit{HST} STIS spectrum of SBS 0909+532 $A$. The cyan and purple curves show the effects of different $b$-values on the fitted Voigt profiles profiles resulting from $b$ = 12 km s$^{-1}$ and $b$ = 45 km s$^{-1}$ respectively. Right panel: The Voigt profile fit for \ion{H}{1} Ly$\alpha$ corresponding to log $N_\mathrm{H\;I}$ = 20.38 $\pm$ 0.20 cm$^{-2}$ in the \textit{HST} STIS spectrum of SBS 0909+532 $B$. The solid colored vertical lines (magenta: \ion{Si}{2} $\lambda$1193; cyan: \ion{N}{1} $\lambda\lambda$1199.6, 1200.2, 1200.7; orange: \ion{Si}{3} $\lambda$1206) indicate the positions (all at $z = \edit1{0.830698}$) of metal lines that may be blended on the left wing of the \ion{H}{1} feature. We were unable to detect \ion{H}{1} Ly$\beta$ toward sight line SBS 0909+532 $B$ because of extreme noise in the region.}
\label{fig:hiAB}}
\end{figure*}

\begin{figure}[htb!]
\centering
\epsscale{1.15}
\plotone{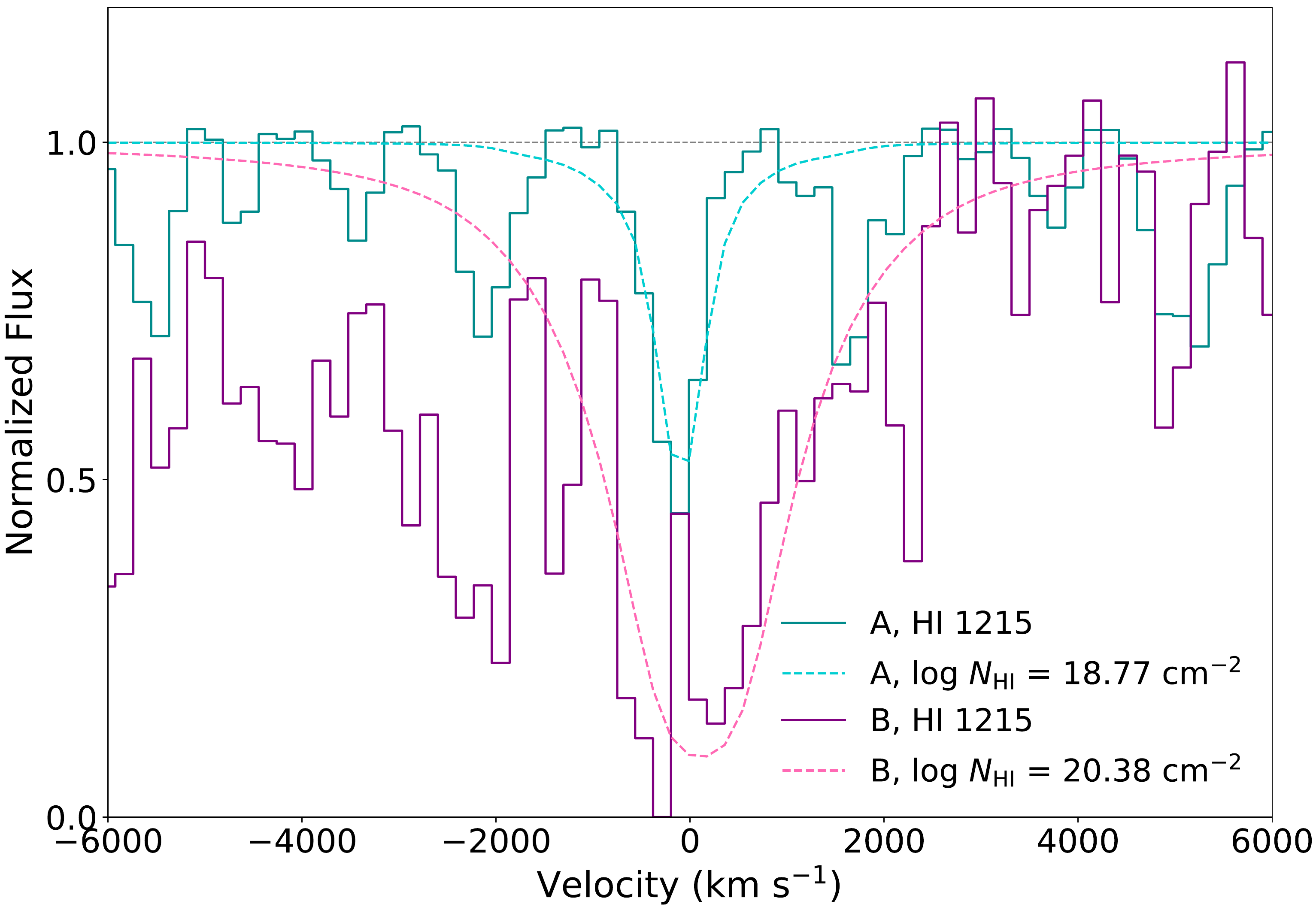}
\caption{Velocity overplot of \ion{H}{1} Ly$\alpha$ in the sight line toward SBS 0909+532 $A$ (gray+blue) with the same feature observed in the sight line toward SBS 0909+532 $B$ (purple+pink).
\label{fig:hiab_oplot}}
\end{figure}

There is significantly more \ion{H}{1} absorption observed in \edit1{the fainter} sight line \textit{B} than in sight line \textit{A}. Given the low resolution of the G230LB data and the broad Ly$\alpha$ feature (see Figures \ref{fig:hiAB} and \ref{fig:hiab_oplot}) both the $z$ and $b$-value for the single \ion{H}{1} component was fixed to the redshift and $b$-value of the dominant metal line component observed in the high-resolution HIRES spectra, and only log $N_\mathrm{H\;I}$ was allowed to vary using \texttt{VPFIT}.

Executing a ${\chi}^2$ minimization analysis to determine damped and sub-damped \ion{H}{1} column densities and uncertainties has historically given rise to unrealistically low uncertainties (see \citealt{2018MNRAS.473.3559P}). Therefore, we created a series of Ly$\alpha$ profiles varying in steps of $\pm$0.1 dex from the resulting Voigt profile fit described in the paragraph above. Comparisons between the profiles were made for the wings and core of the component. For image \textit{B}, it was determined that a $\pm$0.20 dex range from the Voigt profile fit of log~$N_\mathrm{H\;I}$ = 20.38 cm$^{-2}$ gave the most consistency between the fitted profile and the flux within a 2${\sigma}$ buffer. Furthermore, to increase confidence in the adopted log~$N_\mathrm{H\;I}$ and its uncertainty, given the lower signal-to-noise ratio and apparently damped nature of the absorption seen along the sight line to image \textit{B}, we utilized a technique (originally described by \citealt{2000ApJS..130....1R}) to examine the possibility that additional uncertainty may have also arisen from subjective continuum placement. Our original ``most likely'' normalized continuum was shifted above and below by an offset per pixel from the 1${\sigma}$ error array of the flux data. Each offset spectrum was then renormalized and a Voigt profile was fitted again to the \ion{H}{1} Ly$\alpha$ feature. The $+1{\sigma}$ continuum resulted in a column density of log $N_\mathrm{H\;I}$ = $20.11\pm0.20$ cm$^{-2}$. The $-1{\sigma}$ continuum resulted in a column density of log $N_\mathrm{H\;I}$ = $20.55\pm0.25$ cm$^{-2}$. As the mean value of these high and low log $N_\mathrm{H\;I}$ values of $20.38\pm0.33$ cm$^{-2}$ is within the range of the value determined from the initial fit, we chose to retain the initial log~$N_\mathrm{H\;I}$ value of 20.38 cm$^{-2}$ and the conservative $\pm$0.20 dex uncertainty (see Table \ref{tab:lenstabB}). The resulting profile is shown in the right panel of Figure \ref{fig:hiAB}.

\begin{figure*}[htb!]
\epsscale{1.17}
\plottwo{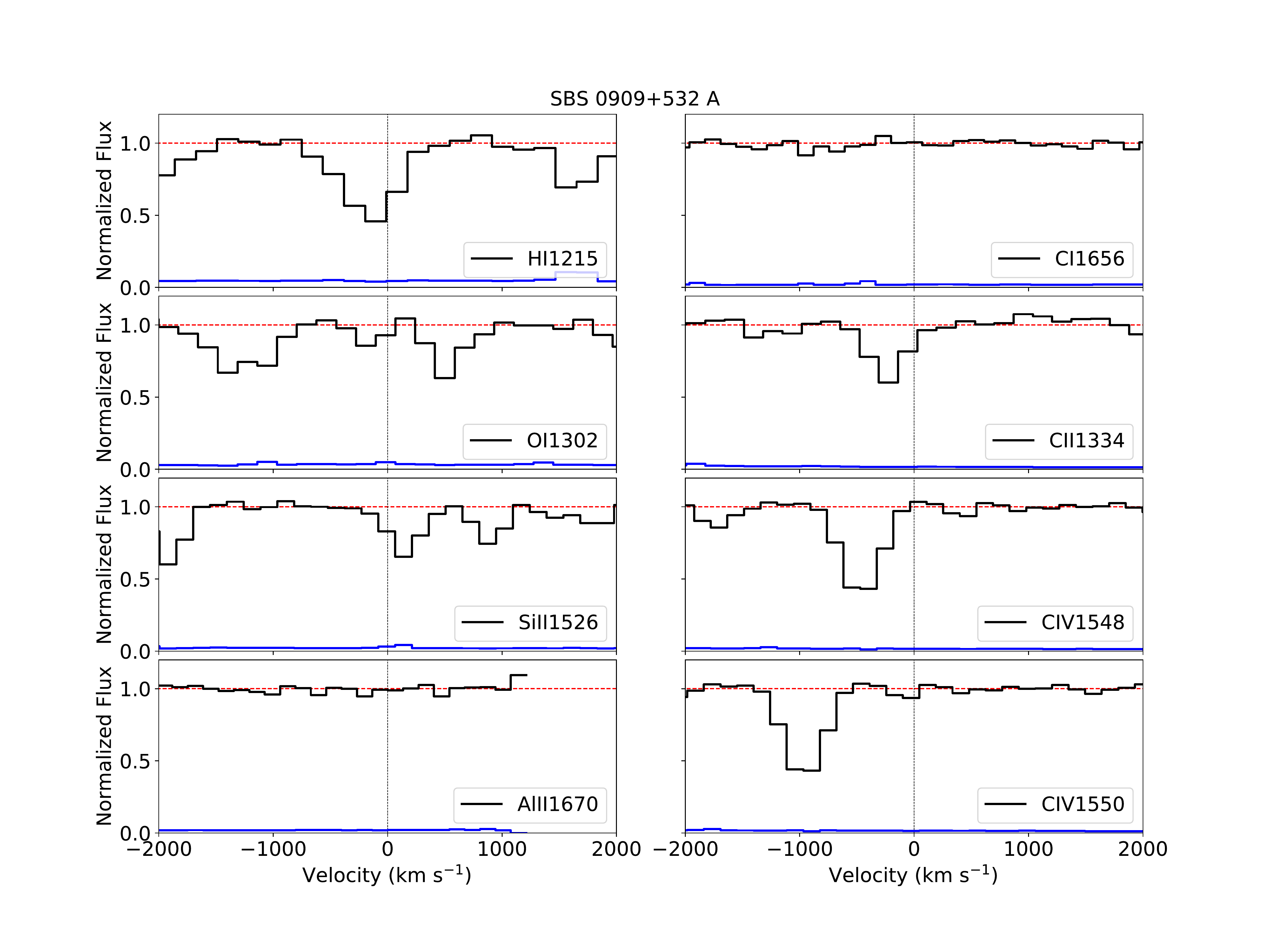}{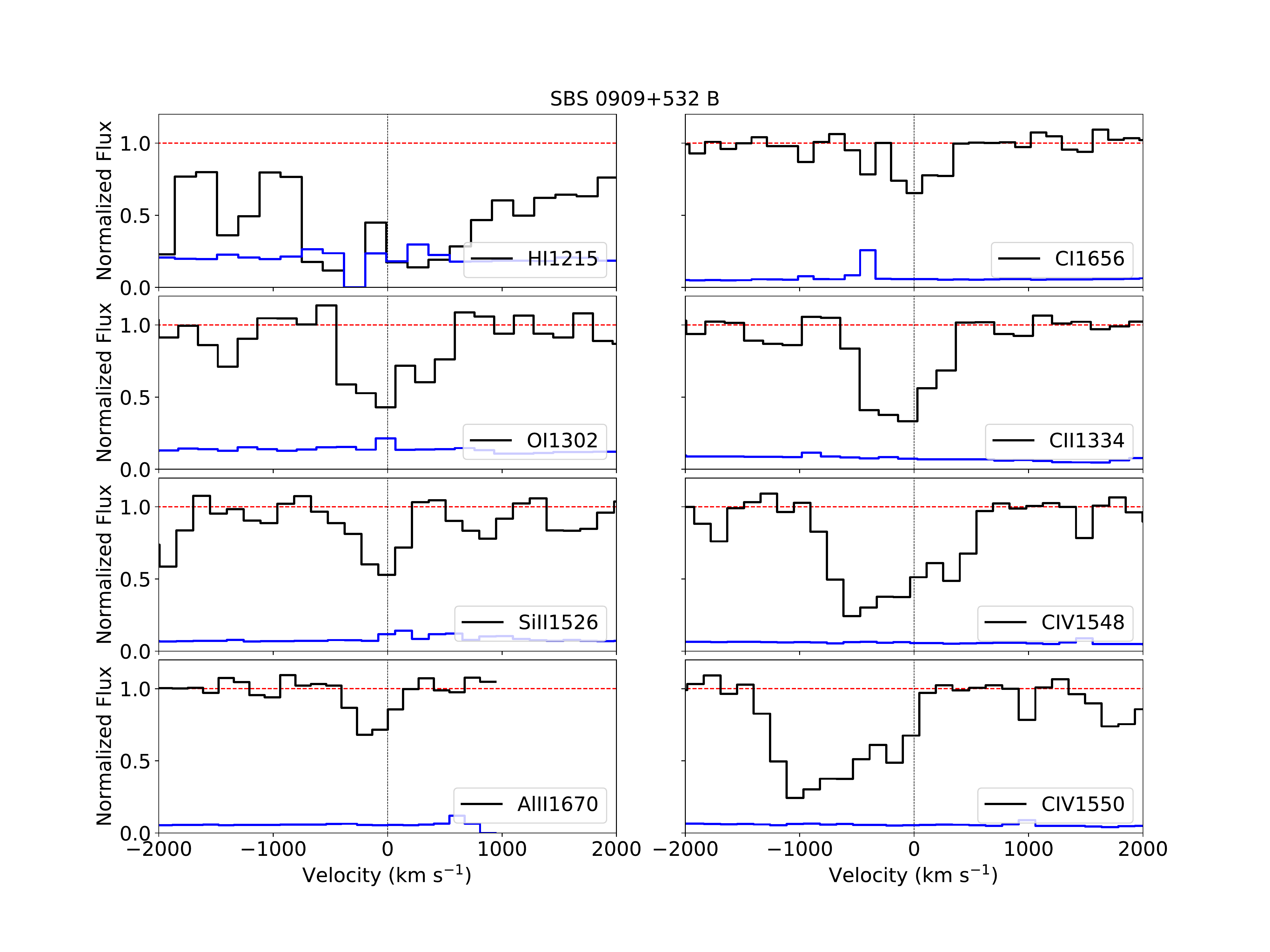}
\caption{Velocity plots of potentially present metal lines with respect to $z$ = 0.83 in STIS image $B$ (right two panels), with the same lines for $A$ (left two panels) included for comparison. There are no data at the red end of the panel for \ion{Al}{2} $\lambda$1670 as this is the end of the available data from the STIS spectrum. Note that the \ion{C}{1} or \ion{Al}{2} transitions may not be present in image \textit{A}.
\label{fig:vplotAB}}
\end{figure*}

The low resolution and signal-to-noise of the STIS spectra are such that column densities of metal species cannot be measured, as their absorption lines are only a few km s$^{-1}$ wide, as shown in the multiple components of the \ion{Mg}{2} and \ion{Fe}{2} lines detected in the corresponding HIRES data (see Figures \ref{fig:metalsA} and \ref{fig:metalsB}). Other potentially existing UV metal lines at the redshift of the lens are shown in Figure \ref{fig:vplotAB}, however, obtaining higher resolution UV spectra is necessary to confirm the identity of these metal lines and to measure the metal column densities. Observations of other metal ions at high resolution would permit more accurate determinations of relative ionization fraction corrections using ion ratios besides the only ratio available to us (\ion{Mg}{2}/\ion{Mg}{1}). Measurements of the undepleted elements S and O would provide more robust determinations of the metallicity, and ion ratios such as \ion{Si}{2}/\ion{S}{2} or \ion{Si}{2}/\ion{O}{1} would allow for dust depletion determinations.

\subsection{Keck HIRES Metal Line Measurements}
\label{subsec:hirescol}
\begin{figure*}[htb!]
\epsscale{0.8}
\plotone{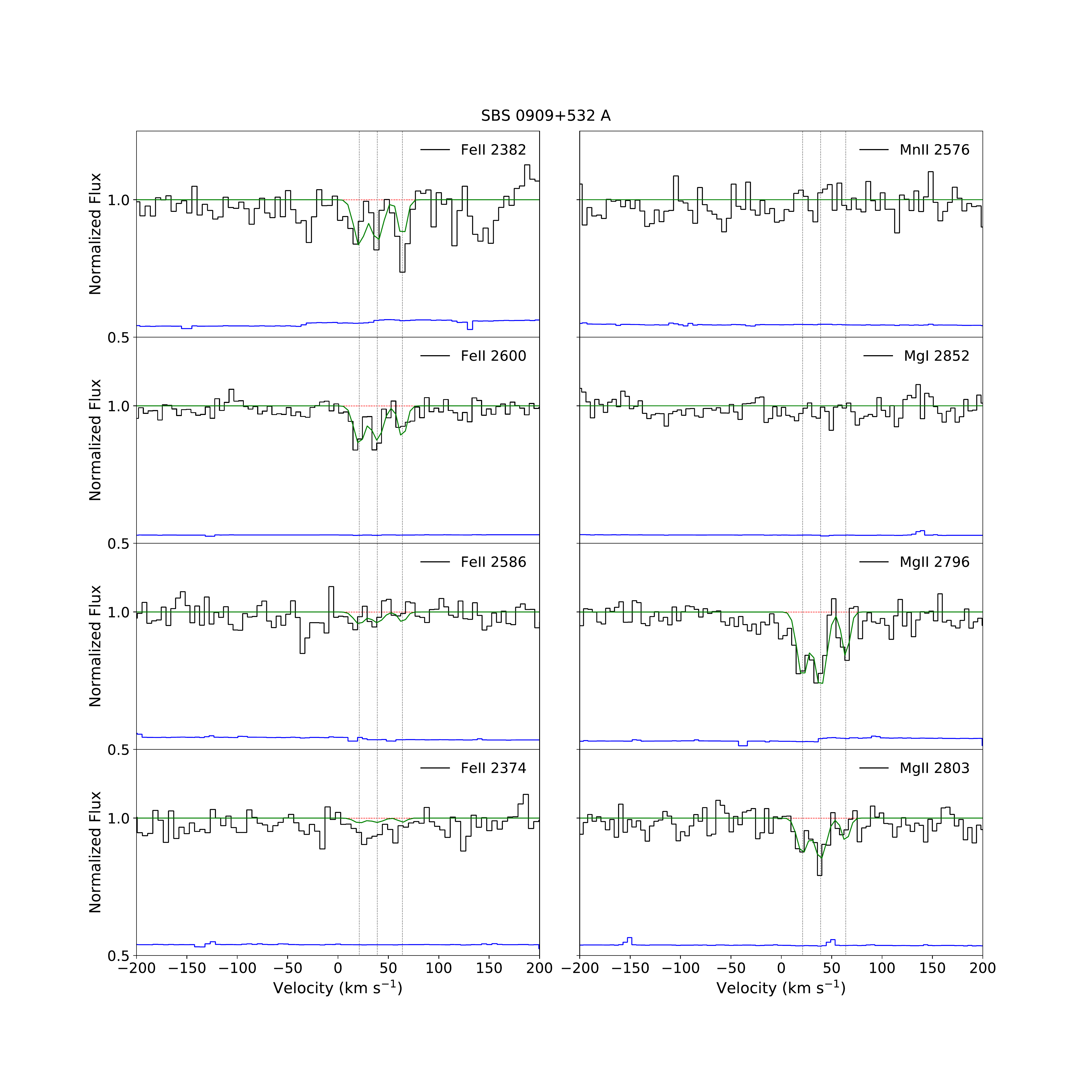}
\caption{Voigt profile fits for the metal lines in the $z \sim0.83$ lens galaxy in the Keck HIRES spectrum of SBS 0909+532 \textit{A}, \edit1{ordered by ion and then by decreasing oscillator strength}. In each panel, the normalized data are shown in black, the solid green curve indicates the theoretical Voigt profile fit to the absorption features, and the dashed red line shows the continuum level. The 1${\sigma}$ error values in the normalized flux are represented by the blue curves near the bottom of each panel. The vertical dotted lines indicate the positions of the components that were used in the fit. As these lines show weaker absorption, the normalized flux scales are shown starting at 0.5 and the 1${\sigma}$ error arrays are offset by 0.5, so that they can be viewed in the same panels. Note that \ion{Mg}{1} \edit1{and \ion{Mn}{2} were} not detected in image \textit{A}, but \ion{Mg}{1} $\lambda2852$ \edit1{and the strongest \ion{Mn}{2} transition at $\lambda2576$ are} included to facilitate comparison with their detection in image \textit{B}, as shown in Figure \ref{fig:metalsB}.
\label{fig:metalsA}}
\end{figure*}

\begin{figure*}[htb!]
\epsscale{0.8}
\plotone{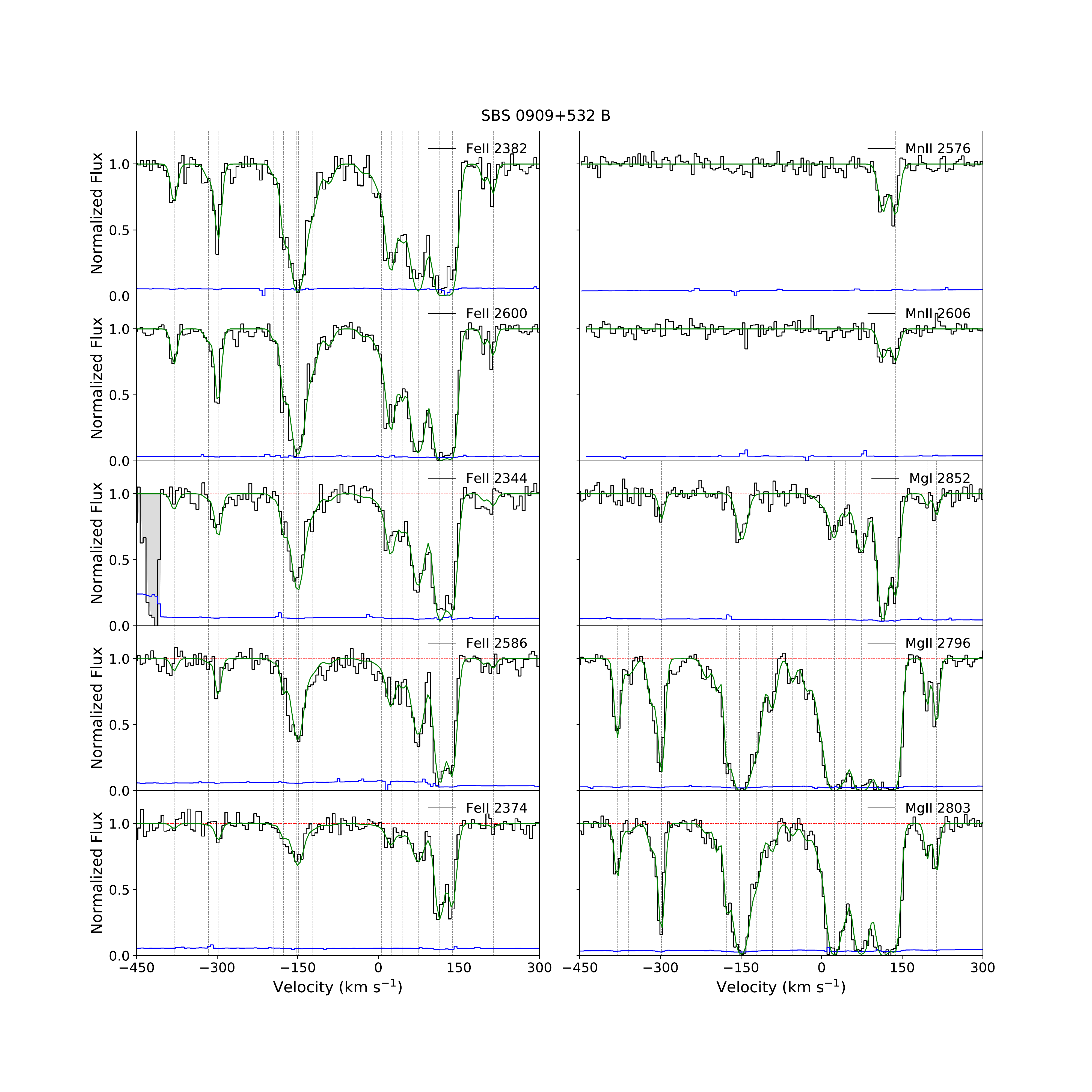}
\caption{Voigt profile fits for the metal lines in the $z \sim0.83$ lens galaxy in the \textit{Keck} HIRES spectrum of SBS 0909+532 \textit{B}, \edit1{ordered by ion and then by decreasing oscillator strength}. In each panel, the normalized data are shown in black, the solid green curve indicates the theoretical Voigt profile fit to the absorption features, and the dashed red line shows the continuum level. The 1${\sigma}$ error values in the normalized flux are represented by the blue curves near the bottom of each panel. The vertical dotted lines indicate the positions of the components that were used in the fit. Shaded regions indicate absorption unrelated to the presented line. 
\label{fig:metalsB}}
\end{figure*}

Tables \ref{tab:lenstabA} and \ref{tab:lenstabB} list individual component column density measurements for the metal line transitions detected in the limited wavelength range of the available HIRES spectra. \edit1{For sight line $A$, all} velocity components seen in \ion{Mg}{2} were also seen in \ion{Fe}{2}. \ion{Mg}{1} \edit1{and \ion{Mn}{2} were} not detected in image \textit{A}; therefore, we calculated the 3${\sigma}$ upper limit to the column density from the 3${\sigma}$ observed frame equivalent width upper limit for \ion{Mg}{1} $\lambda$2852 \edit1{and the strongest \ion{Mn}{2} transition at $\lambda$2576,} assuming a linear curve of growth. We used \texttt{RDGEN} \citep{2014ascl.soft08017C} to select the velocity ranges of the metal lines \edit1{and to initially mark and estimate component column densities. The redshifts of the stronger components were selected by examining the weaker lines, and the redshifts of the weaker components were selected by examining the stronger lines.} For ions where multiple lines were detected, they were fit together to constrain the ionic column densities. The redshifts and $b$-values were also tied together for ions of similar ionization stage. \edit1{Preliminary initial guesses of the redshift, $b$-value, and column density were then fed into \texttt{VPFIT} to determine the final results for the individual components. These parameters were allowed to vary and the program was permitted to add, remove, and/or move the positions of the components until a Voigt profile fit to the data produced the lowest possible ${\chi}^2$.} In image \textit{A}, the absorbing region consists of three very weak components that span a total velocity range of $\sim$66 km s$^{-1}$, \edit1{whereas the absorbing region in image \textit{B} was fit with up to 21 components spanning a total velocity range of $\sim$650 km s$^{-1}$, some of which are saturated. All 21 components were detected in \ion{Mg}{2}, of which 18 could be detected in  \ion{Fe}{2}, 10 in \ion{Mg}{1}, and 2 in \ion{Mn}{2}.} The total column densities for the individual ions are computed by adding all their constituent velocity components together. Voigt profile fits to the spectra are shown in Figures \ref{fig:metalsA} and \ref{fig:metalsB}. The total column densities, as well as the comparable AOD measurements (for the detected lines), can be seen in Table \ref{tab:lenstab}. 

\begin{figure*}[htb!]
\epsscale{0.75}
\plotone{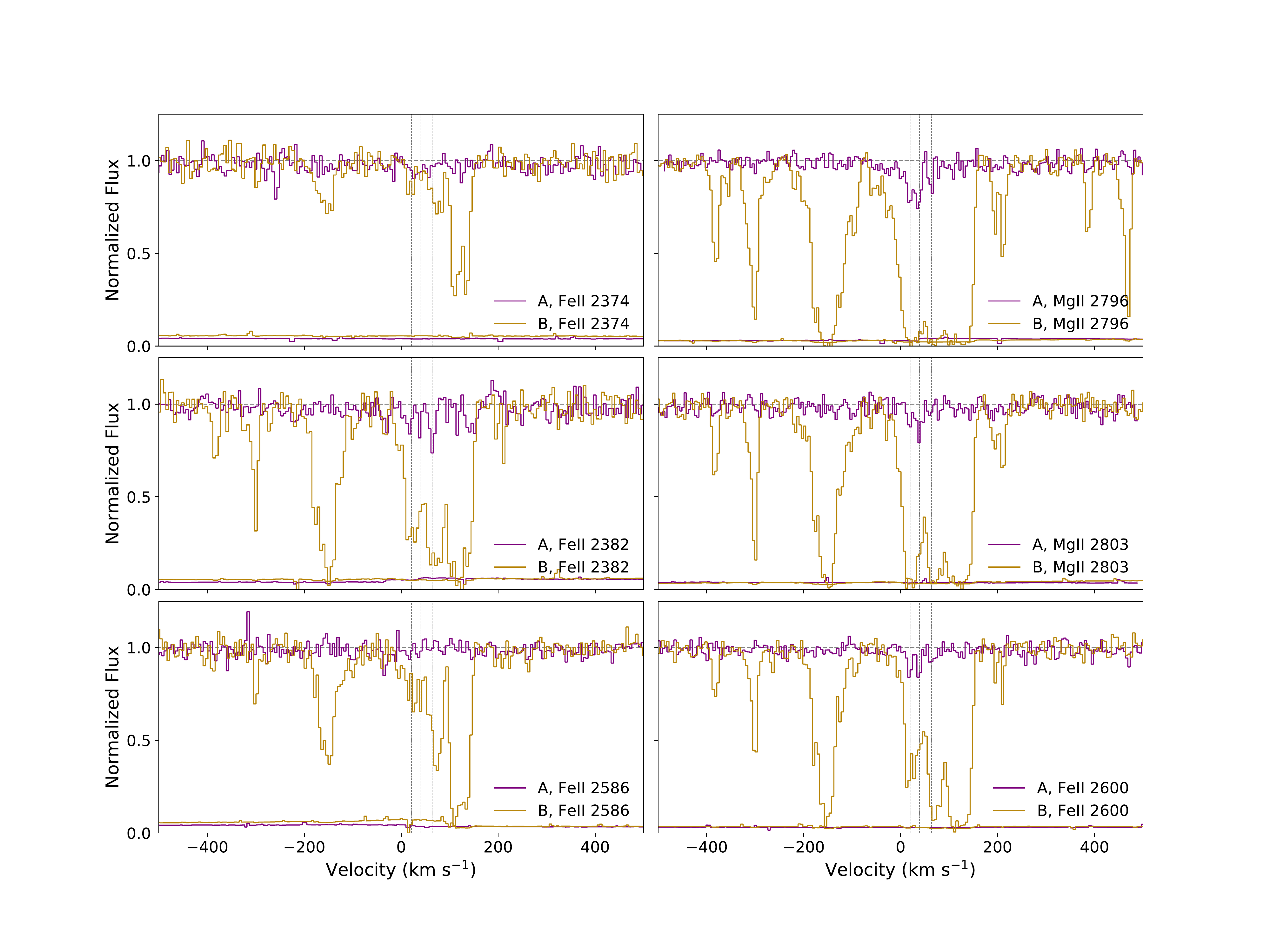}
\caption{Velocity overplots of the detected metal lines in the sight line to the lensed image of SBS 0909+532 $A$ (in purple), with the same metal lines observed in the sight line to the lensed image $B$ (in gold) to illustrate the high degree of difference in absorption between the two lines of sight. The three vertical gray lines at $\sim$21, 39, and 64 km s$^{-1}$ are the locations of the three components in the metal lines observed in the sight line to lensed image $A$. These three components lie entirely within the width of the two components near 28 and 60 km s$^{-1}$ in lensed image $B$.
\label{fig:metal_oplot}}
\end{figure*}

\section{Results} 
\label{sec:results}
The redshifts, $b$-values, and column densities of each sight line's \ion{H}{1} and metal absorption components are summarized in Tables \ref{tab:lenstabA} and \ref{tab:lenstabB}. The total column densities, calculated abundances, and overall average abundance gradients are summarized in Table \ref{tab:lenstab}. Following common practice, the abundances of each element X are defined as [X/H] = log~($N_\mathrm{X}$/$N_\mathrm{H\;I}$) $-$ log~(X/H)$_{\sun}$. Element abundances in the solar photosphere from \cite{2009ARA&A..47..481A} were adopted.

\begin{deluxetable*}{lcccccc}[htb!]
\tabletypesize{\scriptsize}
\tablewidth{0pt} 
\tablecaption{Line parameters in the $z_\mathrm{lens} = 0.83$ galaxy toward SBS 0909+532 \textit{A} \label{tab:lenstabA}}
\tablehead{
\colhead{$z$} & \colhead{$b_\mathrm{eff}$} & \colhead{log $N_\mathrm{H\;I}$\tablenotemark{a}} & \colhead{log $N_\mathrm{Mg\;I}$\tablenotemark{b}} & \colhead{log $N_\mathrm{Mg\;II}$\tablenotemark{c}} & \colhead{log $N_\mathrm{Mn\;II}$\tablenotemark{d}} & \colhead{log $N_\mathrm{Fe\;II}$\tablenotemark{e}} \\
 & (km s$^{-1}$) & (cm$^{-2}$) & (cm$^{-2}$) & (cm$^{-2}$) & (cm$^{-2}$) & (cm$^{-2}$)
}
\startdata
0.830129$\pm$0.000005 & 6.20$\pm$1.11 & ...  & $\leq$ 10.59   & 11.90$\pm$0.04 & $\leq$ 11.58 & 12.08$\pm$0.04 \\
0.830238$\pm$0.000004 & 6.81$\pm$1.01 & ...  & ...                    & 12.00$\pm$0.04 & ...                 & 12.06$\pm$0.07 \\
0.830390$\pm$0.000006 & 3.75$\pm$2.02 & ...  & ...                    & 11.58$\pm$0.09 & ...                 & 11.88$\pm$0.09 \\
\hline
0.829432$\pm$0.000295 & 33.00 & 18.77$\pm$0.12 & ... & ... & ... & ...   \\
\enddata
\tablenotetext{a}{\ion{H}{1} ${\lambda}$1215.6701: $f$-value = 0.4164, \cite{1998PhyS...57..581P}}
\tablenotetext{b}{\ion{Mg}{1} ${\lambda}$2852.964: $f$ = 1.71, \cite{2006ADNDT..92..607F}; There is a non-detection of \ion{Mg}{1} ${\lambda}$2852, thus we calculated the 3${\sigma}$ upper limit to the column density from the 3${\sigma}$ observed frame equivalent width upper limit.}
\tablenotetext{c}{\ion{Mg}{2} ${\lambda\lambda}$2796.352, 2803.531: $f_{2796}$ = 0.613, $f_{2803}$ = 0.306, \cite{2006ADNDT..92..607F}}
\tablenotetext{d}{\ion{Mn}{2} ${\lambda\lambda}$2576.875, 2594.497, 2606.459: $f_{2576}$ = 0.358, $f_{2594}$ = 0.279, $f_{2606}$ = 0.196, \cite{2011ApJS..194...35D}; There is a non-detection of \ion{Mn}{2} ${\lambda\lambda}$2576.875, 2594.497, 2606.459; thus, we calculated the 3${\sigma}$ upper limit to the column density from the 3${\sigma}$ observed frame equivalent width upper limit for the strongest transition at $\lambda$2576.}
\tablenotetext{e}{\ion{Fe}{2} ${\lambda\lambda}$2600.172, 2382.764, 2374.460: $f_{2600}$ = 0.239, $f_{2382}$ = 0.320, $f_{2374}$ = 0.0313, \cite{1996ApJ...464.1044B}; \ion{Fe}{2} ${\lambda}$2586.649: $f_{2586}$ = 0.0717, \cite{2006JPCRD..35.1669F}}
\end{deluxetable*}

\begin{deluxetable*}{lcccccc}[htb!]
\tabletypesize{\scriptsize}
\tablewidth{0pt} 
\tablecaption{Line parameters in the $z_\mathrm{lens} = 0.83$ galaxy toward SBS 0909+532 \textit{B}} \label{tab:lenstabB}
\tablehead{
\colhead{$z$} & \colhead{$b_\mathrm{eff}$} & \colhead{log $N_\mathrm{H\;I}$} & \colhead{log $N_\mathrm{Mg\;I}$} & \colhead{log $N_\mathrm{Mg\;II}$} & \colhead{log $N_\mathrm{Mn\;II}$} & \colhead{log $N_\mathrm{Fe\;II}$} \\
 & (km s$^{-1}$) & (cm$^{-2}$) & (cm$^{-2}$) & (cm$^{-2}$) & (cm$^{-2}$) & (cm$^{-2}$)
}
\startdata
0.827682 $\pm$ 0.000005 & 4.03 $\pm$ 0.59 &  ...  &  ...  & 12.55 $\pm$ 0.03 &  ...  & 12.47 $\pm$ 0.05 \\
0.827811 $\pm$ 0.000017 & 12.52 $\pm$ 4.70 &  ... &  ...  & 11.85 $\pm$ 0.07 &  ...  &  ...  \\
0.828073 $\pm$ 0.000025 & 6.96 $\pm$ 3.25 &  ...  &  ...  & 12.22 $\pm$ 0.03 &  ...  & 12.04 $\pm$ 0.12 \\
0.828183 $\pm$ 0.000004 & 3.54 $\pm$ 1.24 &  ...  & 11.35 $\pm$ 0.10 & 13.57 $\pm$ 0.06 &  ...  & 13.03 $\pm$ 0.04 \\
0.828699 $\pm$ 0.000012 & 9.36 $\pm$ 3.40 &  ...  &  ...  & 11.85 $\pm$ 0.06 &  ...  &  ...  \\
0.828813 $\pm$ 0.000019 & 0.80 $\pm$ 0.73 &  ...  &  ...  & 12.61 $\pm$ 0.21 &  ...  & 11.96 $\pm$ 0.21 \\
0.828922 $\pm$ 0.000007 & 2.91 $\pm$ 4.47 &  ...  &  ...  & 13.04 $\pm$ 0.08 &  ...  & 12.91 $\pm$ 0.06 \\
0.829068 $\pm$ 0.001528 & 13.55 $\pm$ 0.59 &  ... &  ...  & $>$ 13.29 &  ...  & $>$ 13.35 \\
0.829096 $\pm$ 0.000412 & 11.49 $\pm$ 1.92 &  ... & 11.89 $\pm$ 0.04 & $>$ 13.57 &  ...  & $>$ 12.71 \\
0.829258 $\pm$ 0.000015 & 11.11 $\pm$ 0.50 &  ... &  ...  & 12.80 $\pm$ 0.02 &  ...  & 13.46 $\pm$ 0.06 \\
0.829441 $\pm$ 0.000008 & 8.52 $\pm$ 0.74 &  ...  &  ...  & 12.31 $\pm$ 0.03 &  ...  & 12.21 $\pm$ 0.08 \\
0.829672 $\pm$ 0.000009 & 6.54 $\pm$ 2.90 &  ...  &  ...  & 11.85 $\pm$ 0.06 &  ...  &  ...  \\
0.829826 $\pm$ 0.000024 & 7.88 $\pm$ 3.58 &  ...  &  ...  & 11.99 $\pm$ 0.05 &  ...  & 11.79 $\pm$ 0.21 \\
0.830037 $\pm$ 0.000089 & 17.79 $\pm$ 1.48 &  ... & 11.41 $\pm$ 0.12 & $>$ 12.86 &  ...  & 12.60 $\pm$ 0.06 \\
0.830147 $\pm$ 0.000009 & 9.25 $\pm$ 1.28 &  ...  & 11.76 $\pm$ 0.05 & $>$ 13.90 &  ...  & 13.27 $\pm$ 0.03 \\
0.830272 $\pm$ 0.000012 & 3.71 $\pm$ 1.45 &  ...  & 11.22 $\pm$ 0.11 & $>$ 12.77 &  ...  & 12.79 $\pm$ 0.05 \\
0.830453 $\pm$ 0.000003 & 14.74 $\pm$ 1.11 &  ... & 12.11 $\pm$ 0.03 & $>$ 13.79 &  ...  & $>$ 13.71 \\
0.830698 $\pm$ 0.000002 & 8.79 $\pm$ 0.43 & 20.38 $\pm$ 0.20 & $>$ 12.75 & $>$ 14.15 & 12.57 $\pm$ 0.02 & $>$ 14.17 \\
0.830842 $\pm$ 0.000005 & 6.77 $\pm$ 1.90 &  ...  & $>$ 12.32  & $>$ 13.58 & 12.56 $\pm$ 0.02 & $>$ 14.01 \\
0.831199 $\pm$ 0.000003 & 1.65 $\pm$ 0.60 &  ...  & 10.98 $\pm$ 0.19 & 12.40 $\pm$ 0.08 &  ...  & 12.08 $\pm$ 0.13 \\
0.831305 $\pm$ 0.000002 & 2.48 $\pm$ 0.58 &  ...  & 11.27 $\pm$ 0.11 & 12.56 $\pm$ 0.05 &  ...  & 12.34 $\pm$ 0.08 \\
\enddata
\end{deluxetable*}

\subsection{Variations in the \ion{H}{1} Column Density between the Sight Lines}
\label{subsec:hivar}
For the $z_\mathrm{lens}$ = 0.83 absorber toward SBS 0909+532 \textit{AB}, the \ion{H}{1} column density is significantly higher along sight line \textit{B}, by a factor of \edit1{1.61} dex (\edit1{$\sim$41} times higher), indicating that the neutral gas is not distributed homogeneously around the lens. This asymmetry suggests that the sight lines are not probing a spatially coherent region. This difference is interesting for two reasons. First, the separation between the sight lines is small, 1.11${\arcsec}$ or 8.9 kpc at the redshift of the lens. This means that structural differences exist within this normal elliptical galaxy on scales less than 8.9 kpc. Second, the impact parameters ($r$) for sight lines \textit{A} and \textit{B} are $r_A$ = 3.15 kpc and $r_{B}$ = 5.74 kpc; thus sight line \textit{B} is located $\sim$1.8 times further away from the center of the lensing galaxy than sight line \textit{A}. Thus, more \ion{H}{1} exists further from the center of the galaxy, which could suggest that the region probed by sight line \textit{A} at $\sim$3 kpc from the center of the galaxy is highly ionized. SBS 0909+532 \textit{AB} was observed on 2006 December 17 with the Advanced CCD Imaging Spectrometer on board the {\it Chandra} X-ray Observatory by \cite{2009ApJ...692..677D}. Similarly to what we observe, \cite{2009ApJ...692..677D} also saw much more absorption in image \textit{B} than in image \textit{A}. They measured log $\Delta N_\mathrm{H,B-A}$ = 20.74$^{+0.44}_{-0.22}$ dex between the sight lines. \edit1{This difference is consistent with our reported \ion{H}{1} column density measurement for sight line $B$ within the margin of error.} We consider the impact of ionization effects on our measurements for sight line \textit{A} in \S \ref{subsec:cloudy}.

\subsection{Element Abundances and Abundance Gradients}
\label{subsec:abund}
Similarly, we see significantly more metal absorption in sight line \textit{B} than in \textit{A} (see Figure \ref{fig:metal_oplot}). All \ion{Mg}{2} and \ion{Fe}{2} velocity components seen in sight line $A$ are also seen in sight line $B$. However, sight line $B$ shows several velocity components that are not seen in sight line $A$. 

The metallicities were calculated for each sight line and are shown in Table \ref{tab:lenstab}. The average abundance gradient is calculated from the difference in Fe abundances measured in the lensed images and the difference in the impact parameters as measured from the center of the lensing galaxy, i.e., ${\Delta}$[Fe/H]/${\Delta}r$~=~ ([Fe/H]$_B$~$-$~[Fe/H]$_A$)/($r_B$~$-$~$r_A$). This calculation shows how the abundance between the images would change per unit distance if the lens were considered to be uniform. Although Fe depletes readily onto dust grains, we use [Fe/H] to characterize the average abundance gradient because measurements of [Fe/H] exist for other lenses for comparison. Table \ref{tab:lenstab} shows these results. \edit1{The average abundance gradient $\Delta$[Fe/H]/$\Delta~r$~$\geq$~$+$0.20 dex kpc$^{-1}$ between the two sight lines, where the lower limit results from the lower limit of [Fe/H]$_{B}$ from the many saturated components at 150 km s$^{-1}$ $\lesssim v \lesssim -150$ km s$^{-1}$. This positive gradient is much higher} than the range of metallicity gradients observed in the MW and nearby galaxies ($\sim$ $-$0.01 to $-$0.09 dex kpc$^{-1}$ in the MW, \citealt{2002AJ....124.2693F}; \citealt{2011AJ....142..136L}; \citealt{2012ApJ...746..149C};  $-$0.043 dex kpc$^{-1}$ in M101, \citealt{2003ApJ...591..801K};  $-$0.027 $\pm$ 0.012 dex kpc$^{-1}$ in M33, \citealt{2008ApJ...675.1213R}; $-$0.041 $\pm$ 0.009 dex kpc$^{-1}$ in nearby isolated spirals, \citealt{2010ApJ...723.1255R}).

\begin{deluxetable*}{l|ccc|ccc|c}[htb!]
\tabletypesize{\scriptsize}
\tablewidth{0pt} 
\tablecaption{Total Column Densities, Metallicities, and Gradients in the $z_\mathrm{lens} = 0.83$ galaxy \label{tab:lenstab}}
\tablehead{
\colhead{} & \multicolumn{3}{c}{SBS 0909+532 A} & \multicolumn{3}{c}{SBS 0909+532 B} & \colhead{} \\
\cline{2-4} \cline{5-7}
\colhead{Ion} & \colhead{log $N_\mathrm{AOD}$} & \colhead{log $N_\mathrm{fit}$} & \colhead{[X/H]\tablenotemark{a}} & \colhead{log $N_\mathrm{AOD}$} & \colhead{log $N_\mathrm{fit}$} & \colhead{[X/H]\tablenotemark{a}} & \colhead{${\Delta}$[X/H]/${\Delta}r$\tablenotemark{b}}
}
\startdata
Mg {\sc i} & $\leq$ 10.59\tablenotemark{c}        & ... & ... & 13.04$\pm$0.12 & 13.05$\pm$0.02 & ... & ... \\
Mg {\sc ii} & 12.36$\pm$0.08 & 12.34$\pm$0.03 & $-$2.03$\pm$0.12 & 14.35$\pm$0.09 & $\geq$ 14.67 & $\geq$ $-$1.31 &  ... \\
Mn {\sc ii} & $\leq$ 11.58\tablenotemark{d} & ... & $\leq$ $-$0.63 & 12.88$\pm$0.05 & 12.87$\pm$0.02 & $-$0.94$\pm$0.20 & ... \\
Fe {\sc ii} & 12.51$\pm$0.12  & 12.49$\pm$0.04 & $-$1.78$\pm$0.13 & 14.55$\pm$0.24 & $\geq$ 14.62 & $\geq$ $-$1.26 & $\geq$ $+$0.20  \\
\enddata
\tablenotetext{a}{For sight line \textit{B}, [Mg/H] is calculated from the total sum of log $N_\mathrm{Mg\;II}$ and log $N_\mathrm{Mg\;I}$. For sight line \textit{A}, [Mg/H] is calculated from log $N_\mathrm{Mg\;II}$ only due to the non-detection of \ion{Mg}{1}.}
\tablenotetext{b}{Average abundance gradient ${\Delta}$[Fe/H]/${\Delta}r$~=~ ([Fe/H]$_B$~$-$~[Fe/H]$_A$)/($r_B$~$-$~$r_A$) in dex kpc$^{-1}$; does not include ionization corrections.}
\tablenotetext{c}{Non-detection of \ion{Mg}{1} in the lensed image of $A$, column density in cm$^{-2}$ is 3$\sigma$ upper limit to the column density that was calculated based on the 3$\sigma$ observed frame equivalent width upper limit assuming a linear curve of growth.}
\tablenotetext{d}{Non-detection of \ion{Mn}{2} in the lensed image of $A$, column density in cm$^{-2}$ 3${\sigma}$ upper limit to the column density from the 3${\sigma}$ observed frame equivalent width upper limit for the strongest transition at $\lambda$2576.}
\end{deluxetable*}

\subsection{Ionization Effects} 
\label{subsec:cloudy}
\edit1{The metallicities calculated for each sight line made the assumption that the ion stages \ion{H}{1}, \ion{Fe}{2}, and \ion{Mg}{2} can represent the total column density of that element in the Lyman limit system (LLS; sight line \textit{A}) and the damped Ly$\alpha$ absorber (DLA; sight line \textit{B}). The lower $N_\mathrm{H\;I}$ detected in sight line \textit{A} could be an indication that the environment is highly ionized; therefore, we investigate to what extent our results may be affected by ionization of the absorbing gas. In the case of the DLA in sight line $B$, \ion{H}{1} is expected to be self-shielding against photons capable of ionizing it ($h\nu >$ 13.6 eV) as is commonly assumed for DLAs. However, it is necessary to confirm that \ion{Mg}{2} and \ion{Fe}{2} are indeed the dominant ionization stages since \ion{Mg}{1} and \ion{Fe}{1} can be ionized by even photons that cannot ionize \ion{H}{1}, and in principle some of the photons ionizing \ion{H}{1} (those with $h\nu >$ 15.0 and 16.2 eV, respectively) can ionize even \ion{Mg}{2} and \ion{Fe}{2}.} Unfortunately, no higher ions have confirmed detections in either sight line. 

We ran a suite of \texttt{CLOUDY} photoionization models using version 17.01 (\citealt{2017RMxAA..53..385F}) to determine the extent of ionization effects in both sight lines. We used the approximation that the absorption regions are plane-parallel slabs and included the cosmic microwave background at the redshift of the absorber and the extragalactic UV background from \cite{2019MNRAS.484.4174K} at the redshift of the lens (KS18 in \texttt{CLOUDY}) as the radiation fields. \edit1{Additionally, we include the cosmic ray background from \cite{2007ApJ...671.1736I} since cosmic rays not only heat ionized gas but also heat neutral gas and create secondary ionizations.} The neutral hydrogen column densities were  fixed to the estimated values measured from the STIS spectra listed in \edit1{Tables \ref{tab:lenstabA} and \ref{tab:lenstabB}}. The gas metallicities were fixed to the values obtained for Fe from the HIRES spectra (Z$_\mathrm{A,LLS}$ \edit1{$\sim$ 0.02} Z$_{\sun}$ and Z$_\mathrm{B,DLA}$ \edit1{$\gtrsim$ 0.06} Z$_{\sun}$ for sight line \textit{A} and sight line \textit{B} respectively). 

\begin{figure}[htb!]
\epsscale{1}
\plotone{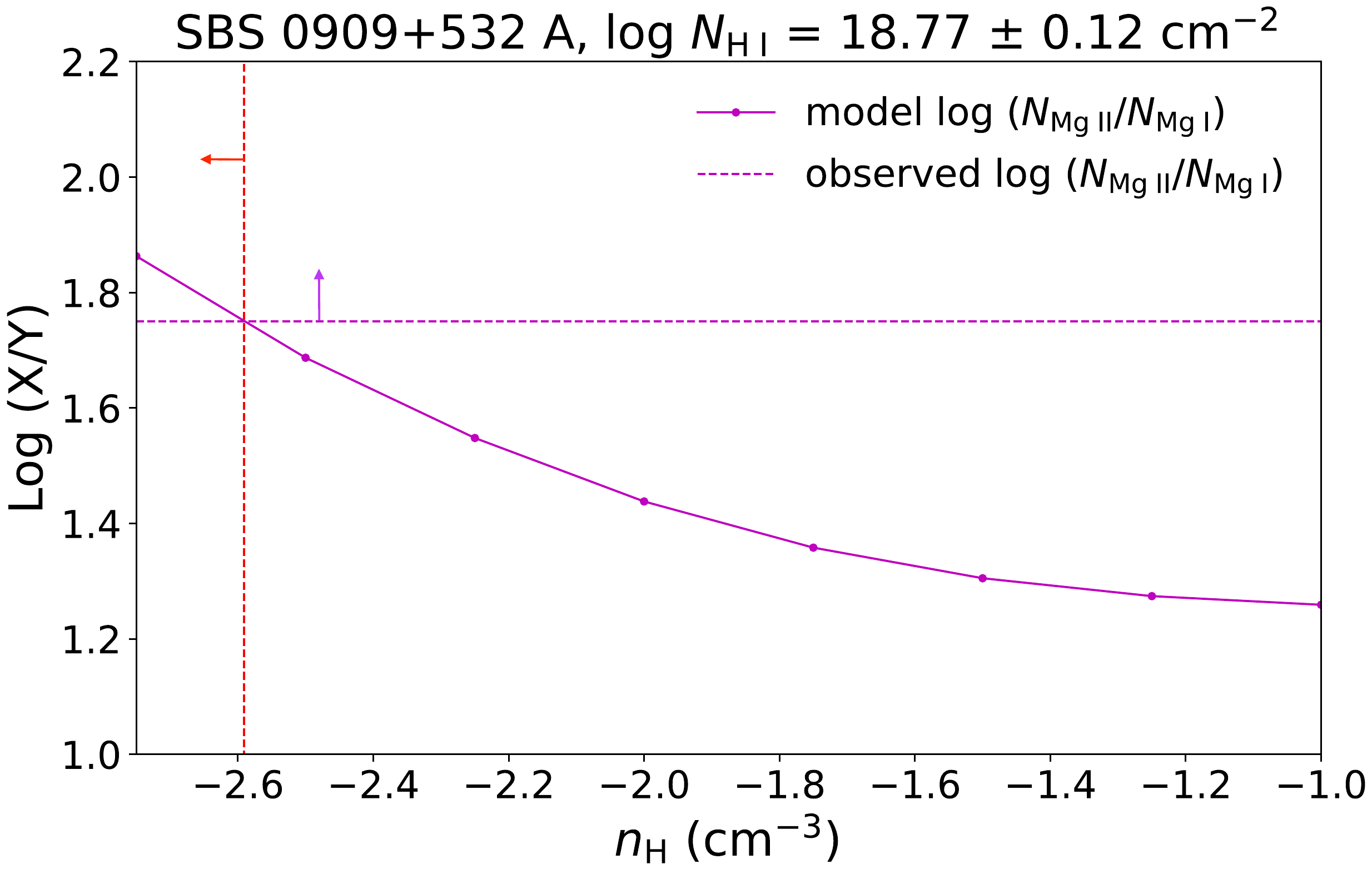}
\epsscale{0.98}
\plotone{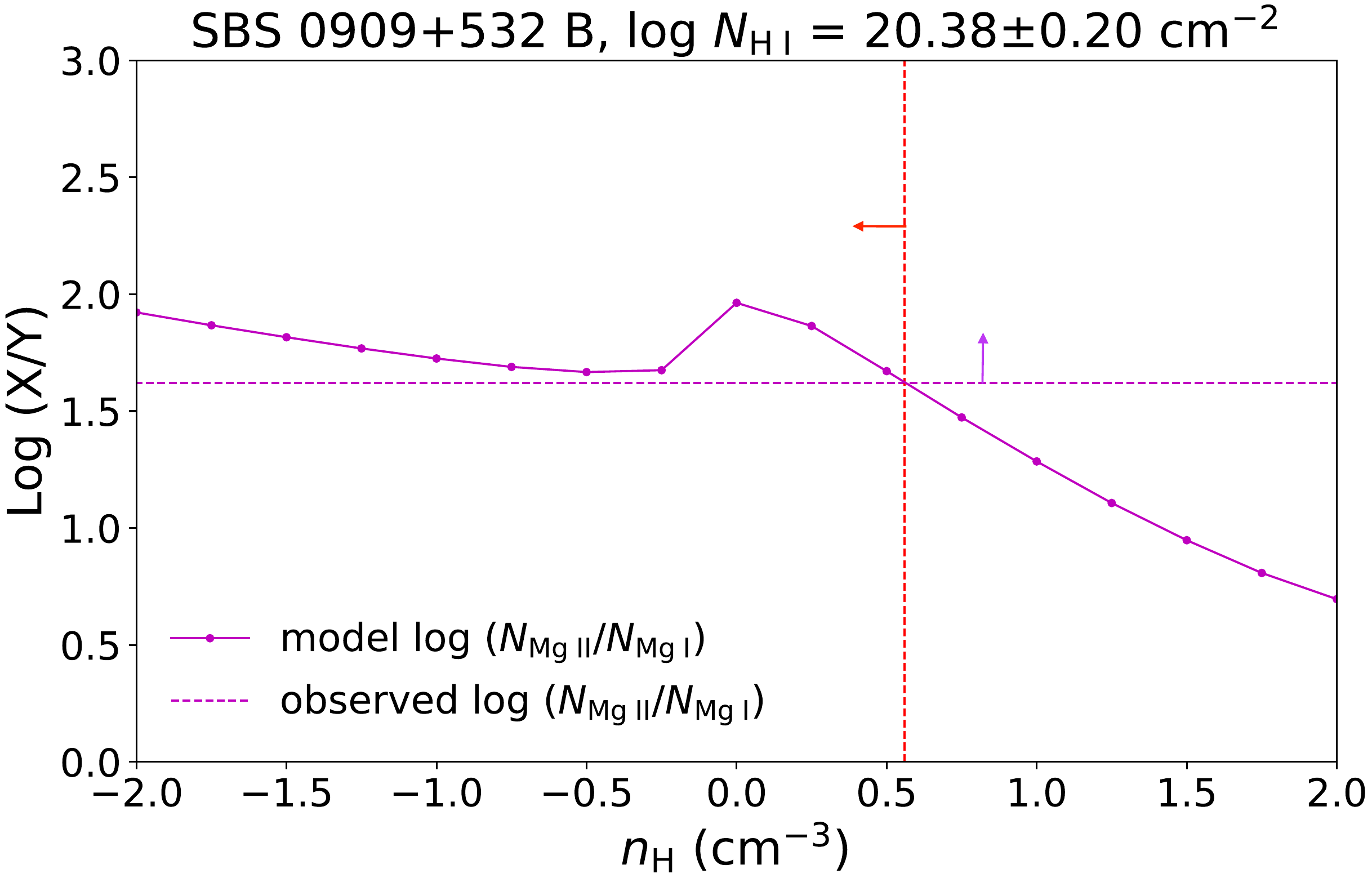}
\caption{Estimated number densities for the two sight lines toward SBS 0909+532 \textit{AB} using \texttt{CLOUDY} version 17.01. \textit{Top}: Comparison of model log $(N_\mathrm{Mg\;II} / N_\mathrm{Mg\;I}$) over a range of log $n_\mathrm{H}$ values and the observed lower limit of log $(N_\mathrm{Mg\;II} / N_\mathrm{Mg\;I})$ $\geq$ 1.75 for the LLS (log $N_\mathrm{H\;I}$ \edit1{= 18.77 $\pm$ 0.12} cm$^{-2}$) observed in the sight line to the lensed image SBS 0909+532 \textit{A}. We estimate log $n_\mathrm{H} \leq$ $-$2.59 cm$^{-3}$ and  log $U \geq -2.56$. \textit{Bottom}: Comparison of model log $(N_\mathrm{Mg\;II} / N_\mathrm{Mg\;I}$) over a range of log $n_\mathrm{H}$ values and the observed log ($N_\mathrm{Mg\;II} / N_\mathrm{Mg\;I}$) \edit1{$\geq$ 1.61} for the DLA (log $N_\mathrm{H\;I}$ = 20.38 $\pm$ 0.20 cm$^{-2}$) at $z = 0.83$ in the sight line to the lensed image of SBS 0909+532 \textit{B}. We estimate log $n_\mathrm{H}$ \edit1{$\leq$ 0.56} cm$^{-3}$ and log $U$ \edit1{$\geq$ $-$5.71}. 
\label{fig:IC}}
\end{figure}

The constraints on the number density log $n_\mathrm{H}$ and the ionization parameter log $U$ were estimated by comparing the value of the only observed column density ratio of adjacent ions available, $N_\mathrm{Mg\;II}/N_\mathrm{Mg\;I}$, to the calculated model ratio for a range of hydrogen number densities from $10^{-3}$ to $10^3$ cm$^{-3}$ (see Figure \ref{fig:IC}). \edit1{We acknowledge that this is a rather broad approach to determine the number density, and thus the ionization parameter, in these systems. For sight line $B$, there are 10 components with both \ion{Mg}{1} and \ion{Mg}{2} detected in the HIRES data. The observed log ($N_\mathrm{Mg\;II}/N_\mathrm{Mg\;I}$) values for these components  range from $\sim$1.3 to 2.2, indicating a variety of degrees of ionization. Unfortunately, a robust component-by-component analysis of the ionization state is not possible without the \ion{H}{1} profiles at higher resolution based on higher-order Lyman series lines; thus, we proceed by considering the ratio of the total of all components for \ion{Mg}{1} and \ion{Mg}{2}.} 

\edit1{The model for sight line \textit{A}} estimates log $n_\mathrm{H}$~\edit1{$\leq$~$-2.59$} cm$^{-3}$ and log $U$ \edit1{$\geq$ $-2.56$}. \edit1{Furthermore, it predicts log $N_\mathrm{H\;II}$ = 20.78 cm$^{-2}$ with total log $N_\mathrm{H}$ = 20.78 cm$^{-2}$, indicating that this region is dominated by ionized hydrogen.} \edit1{For sight line \textit{B}}, log $n_\mathrm{H}$~\edit1{$\leq$ 0.56} cm$^{-3}$ and log $U$ \edit1{$\geq$ $-5.71$}. \edit1{For this cooler sight line, the model estimates log $N_\mathrm{H\;II}$ = 18.29 cm$^{-2}$ and total log $N_\mathrm{H}$ = 20.38 cm$^{-2}$ (essentially equal to log $N_\mathrm{H\;I}$) and thus confirms that neutral hydrogen is the dominant stage. The model also predicts an H$_\mathrm{2}$ column density of 17.57 cm$^{-2}$.} We then estimate the Fe abundance of the gas by correcting the observed column density ratio by the predicted relative ionization fraction for the estimated log $n_\mathrm{H}$, i.e., log (Fe/H) = log ($N_\mathrm{Fe\;II}/N_\mathrm{H\;I}$) $-$ log ($f_\mathrm{Fe^{+}}/f_\mathrm{H^0}$). If the relative ionization fraction $f_\mathrm{Fe^{+}}/f_\mathrm{H^0}$ is $\sim$ 1, then the gas metallicity can be approximated directly from log ($N_\mathrm{Fe\;II}/N_\mathrm{H\;I}$), where then [Fe/H] $\approx$ log ($N_\mathrm{Fe\;II}/N_\mathrm{H\;I}) -$ log (Fe/H)$_{\sun}$.

In sight line \textit{B}, we obtain a relative ionization fraction correction between Fe$^{+}$ and H$^{0}$ of ($f_\mathrm{Fe^{+}}/f_\mathrm{H^0}$) $\approx$ 1 and therefore estimate [Fe/H] directly from log ($N_\mathrm{Fe\;II}/N_\mathrm{H\;I}$) and obtain [Fe/H] \edit1{$\geq$ $-1.26$}. Dividing the model $N_\mathrm{H\;I}$ \edit1{($\approx$ model log $N_\mathrm{H}$)} by model $n_\mathrm{H}$ = \edit1{3.6} cm$^{-3}$, the DLA region is estimated to be \edit1{$\sim$21} pc along the line of sight. 

In sight line \textit{A}, we obtain a relative ionization fraction correction log ($f_\mathrm{Fe^{+}}/f_\mathrm{H^0}$) = \edit1{0.38}. After correcting log ($N_\mathrm{Fe\;II}/N_\mathrm{H\;I}$) by this amount, we adopt \edit1{a metallicity of [Fe/H] = $-2.16$}. This corrected metallicity results in \edit1{an ionization-corrected lower limit of $\Delta$[Fe/H]/$\Delta r$ $\geq$ $+0.35$ dex kpc$^{-1}$. Dividing the model $N_\mathrm{H}$ by the model $n_\mathrm{H}$, the LLS absorbing region is estimated to be $\sim$76 kpc along the line of sight.} 

We note that these models are based on \ion{H}{1} column densities derived from low resolution spectra as well as metal column densities of refractory elements. Higher resolution UV spectra would not only provide more robust \ion{H}{1} column densities, but could also potentially provide additional adjacent ion ratios to better constrain log $n_\mathrm{H}$ and the relative ionization fraction corrections. However, even with a model based on log $N_\mathrm{H\;I}$ measurements from low resolution spectra, the results are consistent with the photoionization studies from \citealt{1986A&A...169....1B}, \edit1{which suggest that  \ion{Mg}{2} absorption dominates in regions of cool, photoionized gas of $T \sim$ 10$^4$ K. We also note that ionization modeling results are sensitive to the atomic parameters used, such as dielectronic recombination coefficients. Improvements to the accuracies of these parameters are essential for improving the reliability of ionization modeling calculations, such as those presented here.}  

\subsection{The Transverse Separation and Mass of the Lens Galaxy}
\label{subsec:mass_sep}
The transverse separation between the GLQ images was calculated for the absorber at $z_\mathrm{abs}$ = 0.611 and the lens at $z_\mathrm{lens}$ = 0.8302. For an absorber with a redshift greater than or equal to the lens redshift, the transverse separation $l_{AB}$ between the sight lines is calculated as
\begin{equation} \label{eqn:lab}
l_{AB} = {D_\mathrm{aq} (1+z_\mathrm{l}) \Delta \theta_{AB} D_\mathrm{l} \over {D_\mathrm{lq} (1+z_\mathrm{a})}}, 
\end{equation}
where $D_\mathrm{l}$, $D_\mathrm{lq}$, and $D_\mathrm{aq}$ are the angular diameter distances between the observer and the lens, between the lens and the quasar, and between the absorber and the quasar, respectively, and $\Delta\theta_{AB}$ is the angular separation between quasar images $A$ and $B$. In the case when the absorber is the lens itself, i.e., $z_\mathrm{l} = z_\mathrm{a}$ and $D_\mathrm{aq}$ = $D_\mathrm{lq}$, and Equation \ref{eqn:lab} simplifies to $l_{AB} = \Delta \theta_{AB} D_\mathrm{l}$. 

The angular diameter distances in Equation \ref{eqn:lab} are calculated using  
\begin{equation} \label{eqn:D}
D_\mathrm{12} = {c \over {H_{0} (1+z_\mathrm{2})}}  \int_{z_\mathrm{1}}^{z_\mathrm{2}} [\Omega_{\Lambda} + \Omega_\mathrm{m}(1+z)^{3}]^{-1/2} \, dz, 
\end{equation}
following \cite{1999astro.ph..5116H}. 

In addition to the unique transverse study of the lens and other absorbers that GLQs provide, analysis of the lensed images provides an opportunity to determine the mass and the mass distribution of the lens. The mass distribution for an early-type galaxy is presumed to be that of a singular isothermal sphere (SIS) given by $\rho \propto r^{-2}$ (e.g., \citealt{2009ApJ...703L..51K}), in which the lens matter behaves as an ideal gas in thermal and hydrostatic equilibrium confined by a spherically symmetric gravitational potential. The velocity dispersion of an SIS of a galaxy lensing a quasar that produces the observed lensed image separation $\Delta\theta$ is 
\begin{equation} \label{eqn:veldisp}
\sigma^2_{SIS} = \frac{c^2 D_\mathrm{q} \Delta\theta}{8\pi D_\mathrm{lq}},
\end{equation}
where $D_\mathrm{q}$ and $D_\mathrm{lq}$ are the same angular diameter distances between the observer and the quasar and the lens and the quasar that were calculated from Equation \ref{eqn:D}. We estimate a velocity dispersion of 258 km s$^{-1}$ for SBS 0909+532. This value is comparable to the velocity dispersion obtained from \cite{1997ApJ...491L...7O} of 272 km s$^{-1}$.

We estimate the mass of the lens galaxy from the astrometry of the lensed images relative to the lens itself using
\begin{equation}
M = -\frac{c^2 \Delta\theta_{AG} \Delta\theta_{BG} D_\mathrm{q} D_\mathrm{l}}{4 G D_\mathrm{lq}},
\end{equation}
where $\Delta\theta_{AG}$ and $\Delta\theta_{BG}$ (with opposite signs) are the angular separations of lensed images $A$ and $B$ from the lens (e.g., \citealt{1992grle.book.....S}). Given the angular separations and our calculated angular diameter distances, we estimate log ($M/M_\sun$) = 11.3. Our value for the mass of the lens of SBS 0909+532 is in agreement with the estimate of log ($M/M_\sun$) = 11.31 by \cite{2000AJ....119..451L}, which they based on the galaxy surface brightness profile within the Einstein radius.

\section{Discussion} 
\label{sec:disc}
Measurements of $N_\mathrm{H\;I}$, and therefore measurements of metallicities, have been performed for only four other lenses (Q1017$-$207 \textit{AB}, Q1355$-$2257 \textit{AB}, \citealt{2019ApJ...886...83K}; HE 0047$-$1756 \textit{AB}, \citealt{2017ApJ...846L..29Z}; HE 0512$-$3329 \textit{AB}, \citealt{2005ApJ...626..767L}) along multiple sight lines  through the lens galaxy. Thus, the measurements for the lens at $z$ = 0.83 toward the two sight lines toward SBS 0909+532 \textit{AB} add important information to this small sample. 

\subsection{\ion{H}{1} Absorption in Lenses} 
\label{subsec:hiabs}
\begin{figure*}[htb!]
\epsscale{1.17}
\plottwo{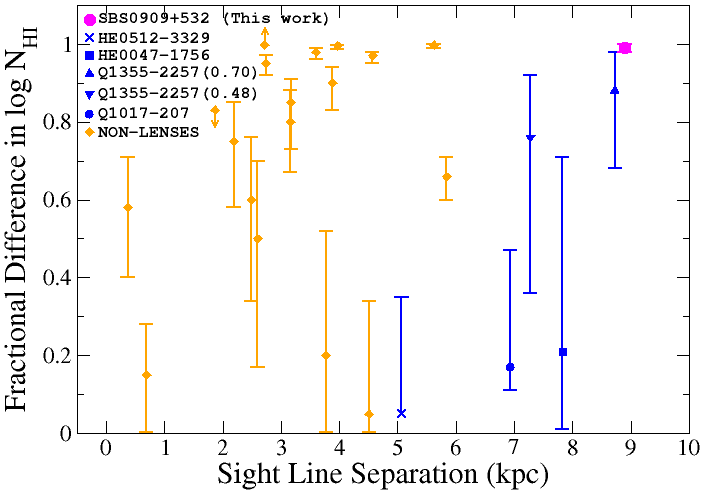}{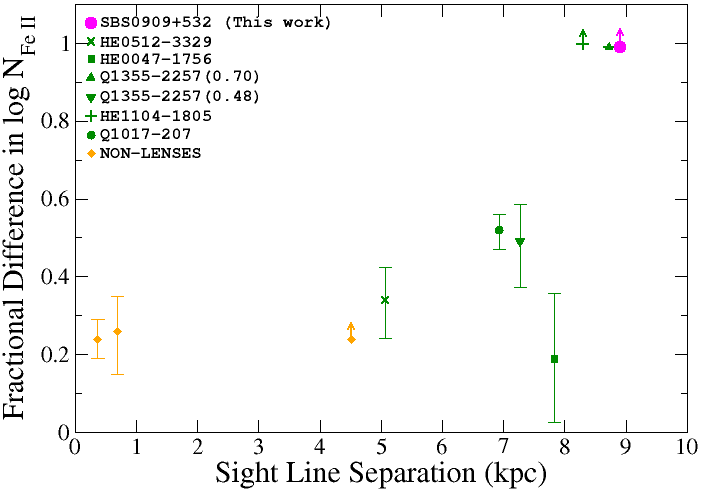}
\caption{Fractional difference in log $N_\mathrm{H\;I}$ (left panel) and log $N_\mathrm{Fe\;II}$ (right panel) for SBS 0909+532 \textit{AB}, calculated as (log $N_\mathrm{X}$ $-$ log $N_\mathrm{Y}$)/log $N_\mathrm{X}$, where sight line X has stronger absorption of the two. The blue and green shapes are lenses for which \ion{H}{1} and \ion{Fe}{2}, respectively, have been measured (Q1017$-$207 \textit{AB}, Q1355$-$2257 \textit{AB}, \citealt{2019ApJ...886...83K}; HE 0047$-$1756 \textit{AB}, \citealt{2017ApJ...846L..29Z}; HE 0512$-$3329 \textit{AB}, \citealt{2005ApJ...626..767L}; HE 1104$-$1805 \textit{AB}, \citealt{2016MNRAS.458.2423Z}, \ion{Fe}{2} only). SBS 0909+532 \textit{AB}, represented with a \edit1{magenta circle}, shows the highest \ion{H}{1} fractional difference between lens sight lines. Only four other lenses to date have measurement of \ion{H}{1} in all sight lines and thus have computed \ion{H}{1} fractional differences; however, five are shown in the figure, as Q1355$-$2257 has two candidates for the lens at $z_\mathrm{lens}$ = 0.48 and 0.70. Regardless, the \ion{H}{1} fractional uncertainty for either lens redshift measurement is in the close range of 0.74$-$0.76, but it differs significantly for \ion{Fe}{2}, shown in the right panel. The orange diamonds in both panels are quasar absorption line systems along the lines of sight to the gravitationally lensed quasars but are not the lenses themselves \edit1{(H1413+117, \citealt{2009MNRAS.397..943M}, \ion{H}{1} only; Q0957+561AB, \citealt{2003ApJ...593..203C}; Q1104$-$1805AB, \citealt{2005ApJ...626..767L}; UM673, \citealt{2010MNRAS.409..679C}, \ion{H}{1} only; SDSS J1442+4055, \citealt{2018A&A...619A.142K}; see Table 26 in \citealt{2019ApJ...886...83K} for a summary of measurements for these particular absorption systems)}. The bars on the points show the maximum and minimum fractional difference possible given the uncertainty in log $N_\mathrm{X}$. 
\label{fig:fracdiffH}}
\end{figure*}

Large differences in \ion{H}{1} and metal column density are observed at small impact parameters on either side of the galaxy. Sight line \textit{A}, with an impact parameter $r_A$ = 3.15 kpc from the lensing galaxy, shows significantly less neutral hydrogen absorption than sight line \textit{B} at an impact parameter $r_{B}$ = 5.74 kpc from the galaxy. The difference between log $N_\mathrm{H\;I,A}$ \edit1{= 18.77 $\pm$ 0.12} cm$^{-2}$ and log $N_\mathrm{H\;I,B}$ = 20.38 $\pm$ 0.20 cm$^{-2}$ of \edit1{1.61} dex shows that the column density of neutral hydrogen drops by a factor of \edit1{41} over a transverse distance of 8.9 kpc. To compare the physical extent of the \ion{H}{1} absorption and the scale over which it varies, we compute the fractional difference in log $N_\mathrm{H\;I}$ measured at the redshift of the lens along both lines of sight, (log $N_\mathrm{H\;I,X}$ $-$ log $N_\mathrm{H\;I,Y}$)/log $N_\mathrm{H\;I,X}$, where sight line X has stronger \ion{H}{1} absorption out of the two, and compare this difference to other lenses in the left panel of Figure \ref{fig:fracdiffH}. In fact, SBS 0909+532 \textit{AB} shows the largest fractional difference in $N_\mathrm{H\;I}$ (\edit1{a difference of 0.98} between log $N_\mathrm{H\;I,B}$ and log $N_\mathrm{H\;I,A}$) for the small sample of lenses for which measurements of $N_\mathrm{H\;I}$ exist. Three other lenses (at $z_\mathrm{lens}$ = 0.408, 0.933, and 1.085 toward quasars HE 0047$-$1756, HE 0512$-$3329, and Q1017$-$2046 respectively) show stronger spatial coherence with a fractional difference in log $N_\mathrm{H\;I} \leq$ 0.20 over separations in the range of 5.07$-$7.83 kpc. The bars on the points show the maximum and minimum possible fractional difference given the range of uncertainty in log $N_\mathrm{H\;I}$. 

While such a large difference in \ion{H}{1} absorption may seem surprising for this small sample of lenses, it is consistent with the $N_\mathrm{H\;I}$  differences seen for other lenses (e.g. Q1355$-$2257 for either $z_\mathrm{lens}$, and even HE 0047$-$1756) within the large error bars. It also may not be uncommon amongst non-lens absorbers (see Figure \ref{fig:fracdiffH}), in which 8 out of the 18 non-lens systems show a fractional difference $>$ 0.90 between sight lines. The orange diamonds are measurements of the fractional difference between $N_\mathrm{H\;I}$ for quasar absorption line systems (LLS, sub-DLAs, and DLAs) along the lines of sight to gravitationally lensed quasars, but they are not the lenses themselves (see Table 26 in \citealt{2019ApJ...886...83K} for details on these absorption systems). However, these non-lens absorbers are likely probing absorption regions belonging to a variety of unknown host galaxies, and it is entirely possible that these lines of sight intersect the galaxy on the same side with larger and similar impact parameters and thus may be expected to show stronger spatial correlations (i.e., \textit{smaller} fractional variations). The large fractional variations seen in some non-lens absorbers could perhaps be explained if those absorbers are associated with less well-mixed gas, e.g., in massive galaxies with more extended star-formation histories. In any case, it is interesting to note that the large difference in \ion{H}{1} (\edit1{1.61} dex) observed between the sight lines of SBS 0909+532 is the highest observed so far amongst the small sample of multiply imaged early-type galaxies. We also note that the large range of possible fractional differences underscores the need for higher resolution UV spectra of these GLQs.

As mentioned in \S \ref{subsec:hivar}, \cite{2009ApJ...692..677D} measured the differential X-ray absorption of the lensing galaxy at $z$ = 0.83 in their effort to study the evolution of the dust-to-gas \edit1{ratio. They} also reported heavier $N_\mathrm{H}$ absorption in image \textit{B} than in image \textit{A} with ${\Delta}$log~$N_\mathrm{H_{B-A}}$ = 20.74$^{+0.44}_{-0.22}$, \edit1{which is consistent with our adopted value of log~$N_\mathrm{H\;I,B}$ = 20.38 $\pm$ 0.20 cm$^{-2}$ within the margin of error.} Hydrogen seen in elliptical galaxies comes in multiple forms, predominately as X-ray-emitting hot gas, perhaps from SNe and stellar winds (e.g., \citealt{1998ApJ...497..681L}, \citealt{2003ARA&A..41..191M}, \citealt{2011A&A...530A..98P}). A large presence of hot hydrogen supports the idea that mature stellar populations could be what prevents reservoirs of chemically enriched cool gas from collapsing into furthering star formation in elliptical galaxies, due to a combination of injected energy from SNe Ia and winds from asymptotic giant branch stars. SBS 0909+532 also happens to reside in a group environment, with three nearby galaxies, two of which are within $\sim$100 kpc, that likely contribute tidal effects (\citealt{2000ApJ...536..584L}). Thus, we cannot rule out that past interactions between group members could have heated or tidally stripped cool gas from the inner regions of SBS 0909+532.  As there is a large difference in both the presence and amount of neutral gas and ions seen between images \textit{A} and \textit{B} given the small impact parameters on either side of the galaxy, it is possible that the galaxy's evolutionary history includes a mixed merger (e.g., a wet-dry merger). 

\subsection{Metal Absorption in Lenses} 
\label{subsec:metabs}
The lack of coherence between the two sight lines separated by 8.9 kpc is clearly seen in the metal absorption lines of the HIRES high resolution spectra in Figures \ref{fig:metalsA} and \ref{fig:metalsB}. We calculated the fractional difference in \ion{Fe}{2} for all lenses in the small sample that also have \ion{H}{1} measurements. These calculations are displayed in the right panel of Figure \ref{fig:fracdiffH}. SBS 0909+532 \textit{AB} \edit1{and HE 1104$-$1805 \textit{AB} \citep{2016MNRAS.458.2423Z} show the highest fractional differences} in \ion{Fe}{2} between lens sight lines in the sample of lenses. In fact, if the galaxy along the line of sight to Q1355$-$2257 at $z$ = 0.70 is not the lens, then SBS 0909+532 \edit1{and HE 1104$-$1805 are} the only lens galaxies in the sample to show a fractional difference in \ion{Fe}{2} above $\sim$0.50. Of course, a relative deficit of Fe II can also be affected by differences in dust depletion, given that Fe is strongly depleted even in the warm MW ISM. We were unable to consider the potential effects due to dust depletion, as no other metal lines were detected.

\subsection{[Fe/Mg] Abundance Ratios} 
\label{subsec:fe/mg} 
As mentioned in \S \ref{sec:intro}, \cite{2016MNRAS.458.2423Z} and \cite{2017ApJ...846L..29Z} investigated [Fe/Mg] ratios in three lens galaxies along sight lines where cool gas was detected to look for possible contributions to the chemical enrichment history of the inner ISM of lenses from SNe Ia. \ion{Mg}{2} absorption traces cool, photoionized gas of $T \sim$ 10$^4$ K (\citealt{1986A&A...169....1B}). Current theories suggest that a combination of injected energy from SNe Ia and winds from asymptotic giant branch stars from mature stellar populations could be responsible for quenching star formation in the reservoirs of chemically enriched cool gas in passive galaxies. \cite{2017ApJ...846L..29Z} reported that while the gas content varied amongst the lenses and within sight lines of the same lenses, supersolar [Fe/Mg] relative abundance patterns were observed in all sight lines that also had detections of cool gas. As each SN Ia is estimated to contribute $\sim$0.7 $M_{\sun}$ of iron and $\lesssim$0.02 $M_{\sun}$ of magnesium (see \citealt{1999ApJS..125..439I}), the high relative abundance they observed suggests a significant contribution to the chemical enrichment of these lens galaxies from SNe Ia. \cite{2017ApJ...846L..29Z} also reported supersolar values of observed log $N_\mathrm{Fe\;II}$/$N_\mathrm{Mg\;II}$ for most of the individual components detected in each sight line toward one of the lens galaxies in their sample, using abundances adopted from \citealt{2006NuPhA.777....1A} (log (Fe/Mg)$_\sun > -$0.08). In both neutral and cool photoionized gas, \ion{Mg}{2} and \ion{Fe}{2} are the respective dominant ionization stages and the ratio $N_\mathrm{Fe\;II}$/$N_\mathrm{Mg\;II}$ should trace the total [Fe/Mg] abundance ratio. Thus, each cloud of cool gas shows the supersolar relative abundance suggestive of a significant contribution from SNe Ia. Another factor that can affect the [Fe/Mg] ratio is that Fe depletes more strongly on dust grains than Mg. This would decrease [Fe/Mg], so the true [Fe/Mg] is even higher. 

\begin{figure}[htb!]
\epsscale{1.1}
\plotone{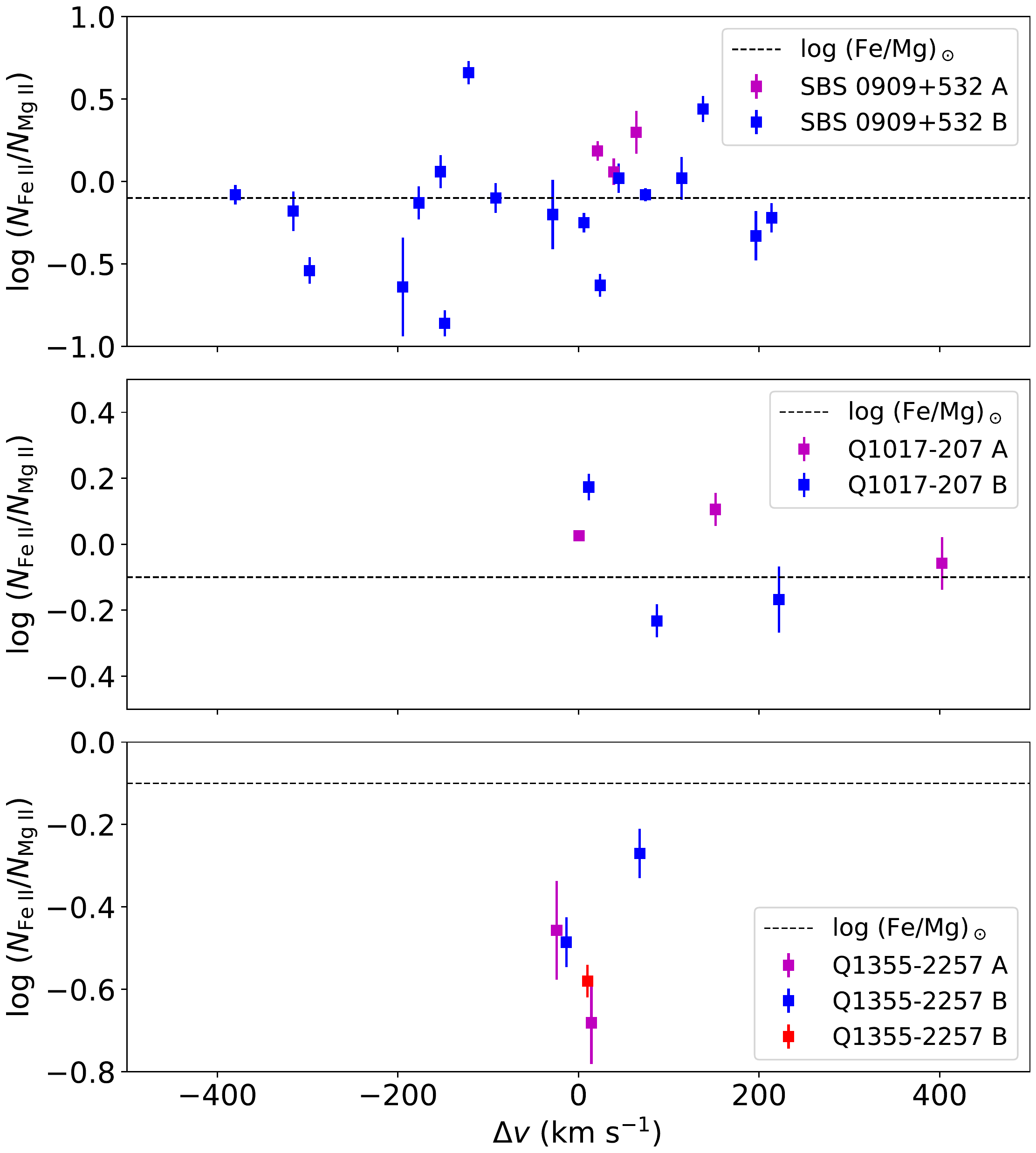}
\caption{Observed column density ratio for log ($N_\mathrm{Fe\;II}/N_\mathrm{Mg\;II}$) for individual components versus their velocity offset from the center of the lens galaxy for SBS 0909+532 \textit{AB}. The black dashed line is the solar ratio log (Fe/Mg)$_{\sun}$ = $-$0.10 from \cite{2009ARA&A..47..481A}. The error bars are the 1${\sigma}$ uncertainties for the calculated ratios. Additionally, we show the same plot for both Q1017$-$207 \textit{AB} and Q1355$-$2257 \textit{AB}, which were observed and reported on in \cite{2019ApJ...886...83K}. \textit{Top}: SBS 0909+532 shows a supersolar value of log ($N_\mathrm{Fe\;II}/N_\mathrm{Mg\;II}$) \edit1{for all components in sight line $A$, but for only 11 out of 18 components in the cooler sight line \textit{B}}. \textit{Middle}: Q1017$-$207 \textit{AB} is doubly imaged by a lens galaxy at $z_\mathrm{lens}$ = 1.0859 with sight lines separated by $l_{AB}$ = 6.9 kpc. Only one of the three resolved components in the MagE spectrum shows a supersolar value for log ($N_\mathrm{Fe\;II}/N_\mathrm{Mg\;II}$) in sight line \textit{A}. All three resolved components show a supersolar value for log ($N_\mathrm{Fe\;II}/N_\mathrm{Mg\;II}$) in sight line \textit{B}. \textit{Bottom}: Q1355$-$2257 \textit{AB} is doubly imaged by a lens galaxy at $z_\mathrm{lens}$ = 0.48 and $l_{AB}$ = 7.3 kpc. None of the components in the MagE spectrum along either sight line show a supersolar value for log ($N_\mathrm{Fe\;II}/N_\mathrm{Mg\;II}$).
\label{fig:femg_vel}}
\end{figure}

The top panel of Figure \ref{fig:femg_vel} shows the observed log $N_\mathrm{Fe\;II}$/$N_\mathrm{Mg\;II}$ values for the individual components resolved in the HIRES spectra of SBS 0909+532 \textit{AB} versus their velocity offset. All three resolved components in sight line \textit{A} show a much higher log $N_\mathrm{Fe\;II}$/$N_\mathrm{Mg\;II}$ value than the typical solar abundance level of $-$0.10 from \cite{2009ARA&A..47..481A}, although low \ion{H}{1} and \ion{Mg}{2} column densities and photoionization modeling suggest that this sight line \edit1{is not} probing cool gas. \edit1{Most of the components in sight line $B$ are saturated and blended. Eleven out of 18 components} detected in sight line \textit{B} show an observed log $N_\mathrm{Fe\;II}$/$N_\mathrm{Mg\;II}$ value higher than the solar value. \edit1{Four of these components can be considered unblended; however, they show subsolar log $N_\mathrm{Fe\;II}$/$N_\mathrm{Mg\;II}$ values. These components are very weak, 11.85~$<$~log~$N_\mathrm{Mg\;II}$~(cm$^{-2}$)~$<$~12.55, and it is possible that these particular components are not tracing cool gas.} The reported effective radius $r_e$ of the lens galaxy is 12.01 $\pm$ 6.84 kpc (\citealt{2000AJ....119..451L}), and therefore the impact parameters of the sight lines are $r_{A} \approx$ 0.26 $r_e$ and $r_{B} \approx$ 0.48 $r_e$. \edit1{Thus, even though there is not a unanimous trend of supersolar relative abundance ratios seen in the cloud components in sight line \textit{B} at this low impact parameter, it does suggest the possibility that the chemical enrichment in this region of the lensing galaxy could be from SNe Ia.}

Additionally, we plot the observed log $N_\mathrm{Fe\;II}$/$N_\mathrm{Mg\;II}$ values for the 
individual components resolved in the MagE spectra of both Q1017$-$207 \textit{AB} and Q1355$-$2257 \textit{AB} (see the middle and bottom panels of Figure \ref{fig:femg_vel}, where both lenses are described in \citealt{2019ApJ...886...83K}). Q1017$-$207 \textit{AB} is doubly imaged by a lens galaxy at $z_\mathrm{lens}$ = 1.086 with sight lines separated by $l_{AB}$ = 6.9 kpc and log $N_\mathrm{H\;I,A}$ = 19.87$\pm$0.09 and log $N_\mathrm{H\;I,B}$ = 19.79$\pm$0.12. We observe supersolar values in three components in the sight line toward Q1017$-$207 \textit{A} and two components in the sight line toward Q1017$-$207 \textit{B}. 1355$-$2257 \textit{AB} is doubly imaged by a lens galaxy at $z_\mathrm{lens}$ = 0.48 and $l_{AB}$ = 7.3 kpc and log $N_\mathrm{H\;I,A}$ = 18.81 $\pm$ 0.18 and log $N_\mathrm{H\;I,B}$ = 19.43 $\pm$ 0.27. The red point shows an alternate value of log $N_\mathrm{Fe\;II}$/$N_\mathrm{Mg\;II}$ if the lens galaxy toward Q1355$-$2257 \textit{B} is instead at redshift $z_\mathrm{lens}$ = 0.70, which is discussed in greater detail in \cite{2019ApJ...886...83K}. None of the components in the sight lines toward Q1355$-$2257 \textit{A} or Q1355$-$2257 \textit{B} at either redshift show a supersolar [Fe/Mg] ratio.

\subsection{Cloud Survival in Early-type Galaxies}
\edit1{The origin of the observed neutral gas in SBS 0909+532 is unknown. We consider the possibility that the absorbing gas may arise in the ISM of the lens galaxy. The gas detected in sight line $A$ shows very low Fe enrichment and if it were ISM dominated, we would expect to see a higher Fe enrichment in the lower impact parameter to the lensed image $A$. For the cooler gas in sight line $B$, we observe $\sim6\%$ solar Fe enrichment, although the true metallicity may be higher since we are currently unable to account for depletion. Our detection of a supersolar [Fe/Mg] relative abundance in 11 out of 18 cloud components could suggest ongoing enrichment in the ISM due to an aging stellar population. For these sight lines positioned on opposite sides of the galaxy, we cannot know to what extent the low impact parameter means a higher probability of hitting the disk, and therefore we consider multiple origin scenarios.}

A recent study by \cite{2019A&A...625A..11A} explores multiple interpretations of the origin and fate of cool circumgalactic gas clouds associated with massive early-type galaxies. Their favored interpretation is that the cool clouds originate as the result of filaments of low metallicity intergalactic gas accreting into the galaxy halo, which then fall in through the halo. They conclude that it is very unlikely that these clouds survive the entire infall and instead evaporate in the hot corona. Thus, the internal regions of the halo are mostly devoid of cool gas, and early-type galaxies are quiescent even though large reservoirs of cool circumgalactic gas are present. On the other hand, \cite{2020MNRAS.498.2391N} report the possibility of harboring cool gas at small impact parameters based on high-resolution TNG50 cosmological magnetohydrodynamical simulations, which explore the origin and properties of cold circumgalactic medium (CGM) gas around massive early-type galaxies at $z$ $\sim$0.5. They find a remarkable population of small-scale, cold gas structure in the CGM of these elliptical systems traced by neutral \ion{H}{1} and \ion{Mg}{2}, concluding that these halos may host $\sim$10$^4$ discrete absorbing cloudlets, approximately a kiloparsec or smaller in size. In this scenario, it is not surprising to find \ion{H}{1} column densities as high as that observed in SBS 0909+532 $B$ at an impact parameter of 5.74 kpc away from the galaxy center.

However, if we consider the neutral gas observed in sight line $B$ within the context of it being of external origin, as described in \cite{2019A&A...625A..11A}, then we can hypothesize possible scenarios for this low metallicity cloud's survival. We do note that the abundance is based on Fe ([Fe/H]$_B$ \edit1{$\geq$ $-1.26$}), which depletes readily into ISM dust; thus, the true metallicity could be higher. If the cloud were indeed metal-poor and originated from the intergalactic medium, one may expect this surviving cloud to be an outlier in order to have survived the infall through the hot corona. Another possibility is that the cloud did not experience fatalistic interactions as it was infalling, which may suggest that SBS 0909+532 has a patchy CGM. Or perhaps the cloud (at a projected distance of 5.74 kpc from the lens galaxy) is not infalling gas originating from the intergalactic medium after all, but is a remnant of a prior interaction with another galaxy. Indeed, it is quite possible that the cool gas cloud arises from local cooling inflows and is dominated by magnetic pressure, given that such gas can in fact exist far inside the virial radius, as seen in the simulations of \cite{2020MNRAS.498.2391N}. The reason it has not collapsed further to continue forming stars could be due to the galaxy's overall aging stellar population, e.g., energy interjection from SNe Ia and/or winds from AGB stars. As mentioned in \S\ref{subsec:fe/mg}, we detect a supersolar [Fe/Mg] relative abundance \edit1{in 11 out of 18 cloud components in the cooler sight line to the lensed image of SBS 0909+532 $B$, which supports a possibility that we are seeing signatures of energy interjection}. 

\subsection{An \ion{Mg}{2} absorber at $z_{abs} \approx$ 0.611}
\label{subsec:abund06}
\begin{figure}[htb!]
\epsscale{0.9}
\plotone{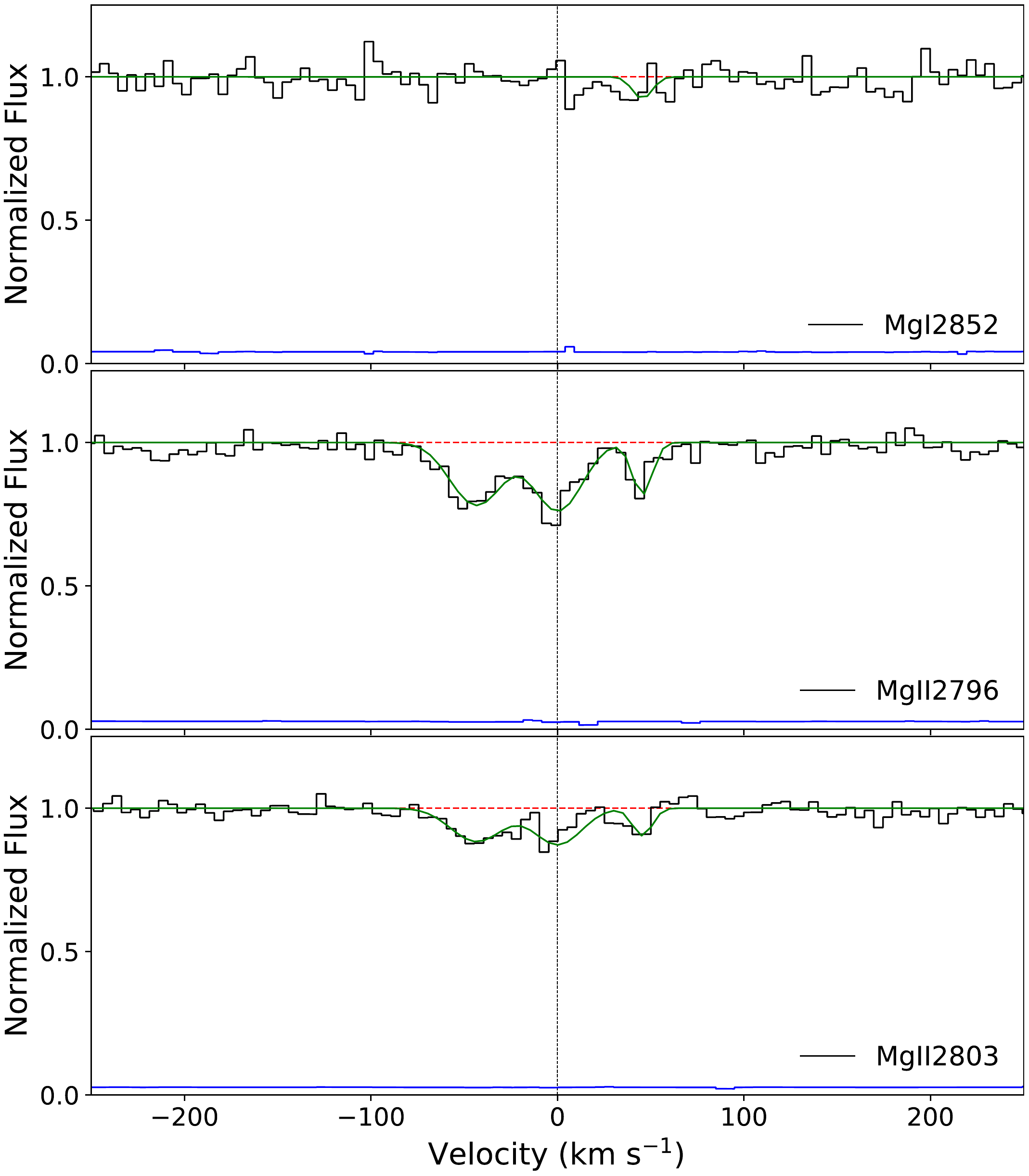}\vspace{3mm}
\plotone{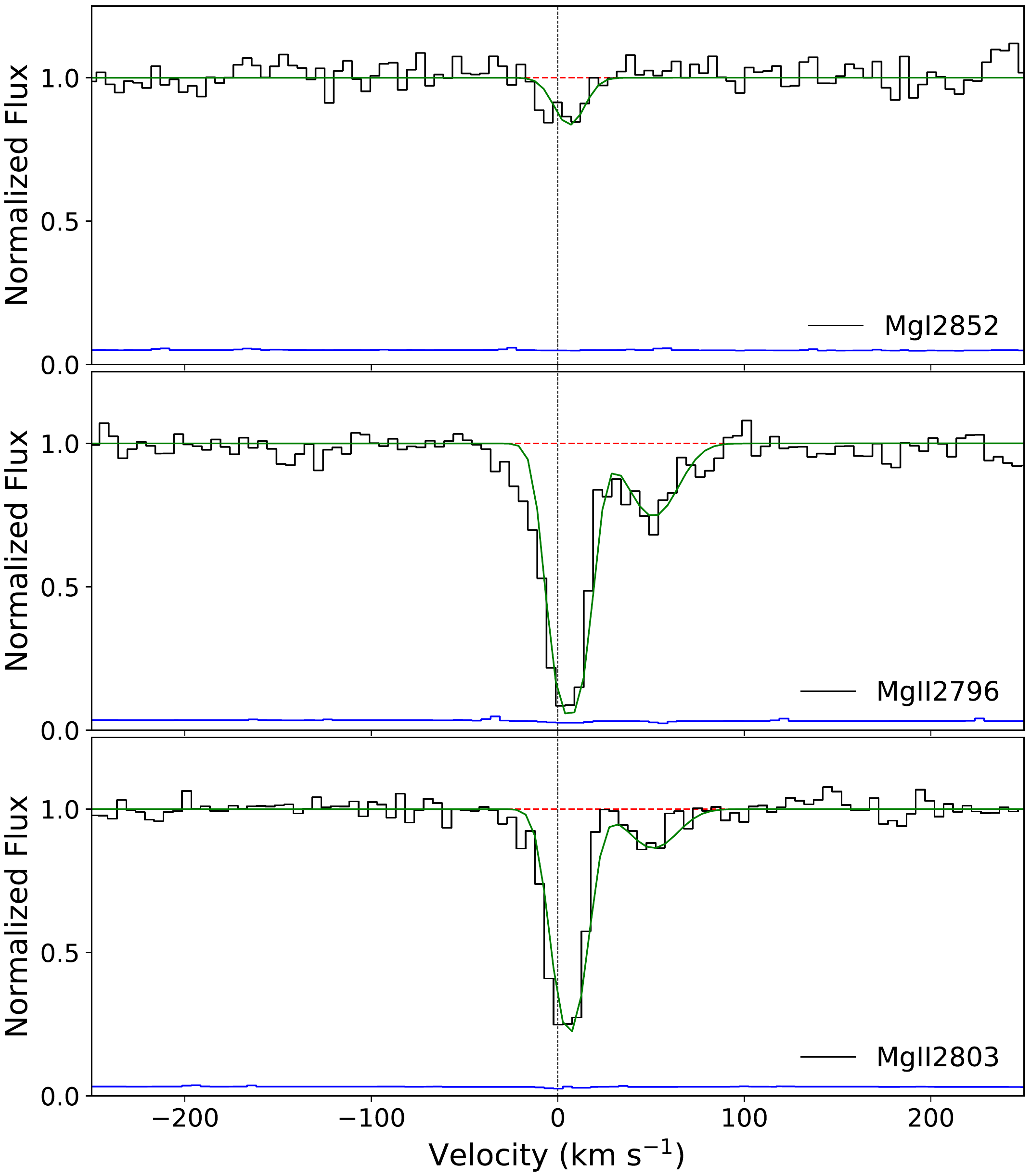}
\caption{Voigt profile fits of the \ion{Mg}{2} absorber detected at $z=0.6116$ toward SBS 0909+532 \textit{A} (top) 
and $z=0.6114$ toward SBS 0909+532B (bottom). In each 
panel, the normalized data are shown in black, the solid green curve indicates the theoretical 
Voigt profile fit to the absorption features, and the dashed red line shows the continuum level. 
The 1${\sigma}$ error values in the normalized flux are represented by the blue curves near the 
bottom of each panel. 
\label{fig:mgabs}}
\end{figure}

While examining the HIRES spectrum of SBS 0909+532 \textit{AB} for metal lines at the redshift of the lens, we found an \ion{Mg}{2} absorption system in both sight lines with the dominant \ion{Mg}{2} component centered at a redshift of $z_\mathrm{abs,A}$ = 0.6116 and $z_\mathrm{abs,B}$ = 0.6114. No other metal lines were detected at these redshifts. The STIS spectrum was examined for corresponding \ion{H}{1} absorption; however, none was seen due to extremely high noise at the blue end. The \ion{Mg}{1} and \ion{Mg}{2} column densities were measured using both Voigt profile fitting and the AOD method where possible. The Voigt profile fits are shown in Figure \ref{fig:mgabs}. The equivalent widths of the \ion{Mg}{2} lines are listed in Table \ref{tab:absorbers} and the column densities for the individual components and the total column densities are listed in Table \ref{tab:mgabsAB}. While the \ion{Mg}{1} and \ion{Mg}{2} absorption is stronger in sight line $B$ compared to sight line $A$, the \ion{Mg}{2}/\ion{Mg}{1} ratio is comparable in both sight lines, with $N_\mathrm{Mg\;II}$/$N_\mathrm{Mg\;I}$ $\sim$53 in sight line $A$ and $N_\mathrm{Mg\;II}$/$N_\mathrm{Mg\;I}$ $\sim$56 in sight line $B$. Although this system is unrelated to the lens galaxy, we report its existence for those interested in lensed \ion{Mg}{2} absorption line systems.

\begin{deluxetable*}{lccc|lccc}[htb!]
\tabletypesize{\scriptsize}
\tablewidth{0pt} 
\tablecaption{Results of Voigt Profile Fitting for Ions in the $z$ = 0.611 \ion{Mg}{2} Absorber along the Sight Line toward SBS 0909+532 $A$ and $B$ \label{tab:mgabsAB}}
\tablehead{
\colhead{} & \multicolumn{3}{c}{SBS 0909+532 $A$} & \colhead{} & \multicolumn{3}{c}{SBS 0909+532 $B$} \\
\cline{2-4} \cline{6-8}
\colhead{$z_A$} & \colhead{$b_\mathrm{eff}$\tablenotemark{a}} & \colhead{log $N_\mathrm{Mg\;I}$\tablenotemark{b}} & \colhead{log $N_\mathrm{Mg\;II}$\tablenotemark{c}} &
\colhead{$z_B$} & \colhead{$b_\mathrm{eff}$\tablenotemark{a}} & \colhead{log $N_\mathrm{Mg\;I}$\tablenotemark{b}} & \colhead{log $N_\mathrm{Mg\;II}$\tablenotemark{c}}
}
\startdata
0.611367$\pm$0.000007 & 18.57$\pm$1.89             & ...                        &12.26$\pm$0.02 & 0.611433$\pm$0.000001 & 10.44$\pm$0.25 & 11.44$\pm$0.09 & 13.13$\pm$0.01 \\
0.611601$\pm$0.000006 & 17.04$\pm$1.48             & ...                        &12.27$\pm$0.02 & 0.611675$\pm$0.000005 & 17.24$\pm$1.51 & ...                        & 12.31$\pm$0.02 \\
0.611845$\pm$0.000004 & 6.19$\pm$1.26               & 10.91$\pm$0.18 &11.77$\pm$0.05 &                                          &                            &                            &                            \\
\hline
Total log $N_\mathrm{X,fit}$   &                                        & 10.91$\pm$0.18 & 12.63$\pm$0.01 & Total log $N_\mathrm{X,fit}$    &                            & 11.44$\pm$0.09 & 13.19$\pm$0.01 \\
Total log $N_\mathrm{X,AOD}$  &                                    &    ...                     & 12.59$\pm$0.04 & Total log $N_\mathrm{X,AOD}$  &                            & 11.61$\pm$0.08 & 13.18$\pm$0.02 \\
\enddata
\tablenotetext{a}{In km s$^{-1}$}
\tablenotetext{b}{Measurements determined from \ion{Mg}{1} ${\lambda}$2852.964 line and are in cm$^{-2}$.}
\tablenotetext{c}{Measurements determined from \ion{Mg}{2} ${\lambda\lambda}$2796.352, 2803.531 lines and are in cm$^{-2}$.}
\end{deluxetable*}

\section{Conclusion} 
\label{sec:conc}
We examined both \textit{HST}-STIS UV spectra and \textit{Keck} HIRES optical spectra of the images of the doubly lensed quasar SBS 0909+532 at $z_\mathrm{QSO}$ = 1.37 to study the spatial differences in neutral hydrogen and metal absorption lines of the lens galaxy at $z_\mathrm{lens}$ = 0.83. We detect a significant difference in \ion{H}{1}, \ion{Mg}{2}, and \ion{Fe}{2} between the lensed images separated by 8.9 kpc. We calculate a large fractional difference \edit1{($\geq$ 0.98)} in the column densities of both neutral hydrogen and metals at impact parameters of $r_A$ = 3.15 kpc (0.26 $r_e$) and $r_B$ = 5.74 kpc (0.48 $r_e$) from the lens galaxy ($r_e \sim$12 kpc). We measure log $N_\mathrm{H\;I,A}$ \edit1{= 18.77 $\pm$ 0.12} cm$^{-2}$ and log $N_\mathrm{H\;I,B}$ = $20.38\pm0.20$ cm$^{-2}$ in the STIS spectra and find that the lens for SBS 0909+532 \textit{AB} shows the highest fractional difference in \ion{H}{1} of lensing galaxies for which only four other \ion{H}{1} measurements currently exist (see the left panel of Figure \ref{fig:fracdiffH}). High ionization in the region probed by sight line \textit{A} is likely responsible for these differences in \ion{H}{1}, as \cite{2009ApJ...692..677D} also reported similar differences in X-ray absorption. The iron abundance is low in both sight lines, [Fe/H] \edit1{= $-1.78$} (uncorrected) in image \textit{A} and [Fe/H] \edit1{$\geq$ $-1.26$} in image \textit{B}, but could be higher possibly due to dust depletions for which we are unable to account, given the limited wavelength range of the optical spectra. Thus, we are unable to ascertain as to what extent dust depletion may be influencing the differences we observe. \edit1{We performed ionization models on both sight lines, with the model for sight line $A$ suggesting that the region is highly ionized. The resulting relative ionization fraction is $f_{Fe+}/f_{H0}$ = 0.38, and we adopt [Fe/H] = $-2.16$} for sight line $A$. \edit1{Ionization corrections were unnecessary for sight line $B$, as the model determined that the relative ionization fraction is $\sim$1.} We also determine \edit1{a lower} limit average abundance gradient \edit1{$\geq$ +0.35} dex kpc$^{-1}$ based on a relative ionization fraction corrected [Fe/H] and a sight line separation of 8.9 kpc, showing that the metallicity \edit1{could be} increasing with radius for this particular lens galaxy. However, we emphasize that higher resolution UV spectra of the individual images are essential to obtain even firmer values on both log $N_\mathrm{H\;I}$ for sight lines \textit{A} and $B$ and to measure column densities of additional metals of higher ionization stages \edit1{in order to explore the multiphase nature of the absorbing gas}. These measurements would permit a more robust determination of metallicities and ionization fraction corrections, which would be beneficial since our initial findings suggest that these significant differences are due to differences in ionization. We also compare our results with those from recent studies suggesting that SNe Ia make significant contributions to the chemical enrichment of the environment of elliptical galaxies (e.g., \citealt{2016MNRAS.458.2423Z}, \citealt{2017ApJ...846L..29Z}, and \citealt{2017A&A...603A..80M}). We find that \edit1{11 of 18 cloud components in the cooler sight line $B$ show} a supersolar $N_\mathrm{Fe\;II}/N_\mathrm{Mg\;II}$ value, which supports the idea that energy interjection from aging stars in elliptical galaxies may prevent reservoirs of cool gas from collapsing to further star formation. Finally, even though the origin of the cold gas detected along the line of sight to image $B$ is unknown, we discuss its existence within the context of some models of cloud survival. We find that our observations resemble the scenario described in the simulations from \cite{2020MNRAS.498.2391N}, who predict a population of small-scale ($<$ 1 kpc) cold gas cloudlets in the CGM of elliptical systems traced by \ion{H}{1} and \ion{Mg}{2}. In this framework, it would not be unexpected for a high \ion{H}{1} column density cloud to survive at an impact parameter of 5.74 kpc from the center of the lens galaxy. 

Robust measurements of both volatile and refractory elements will enable determination of the differences in dust depletions between the different sight lines. These differences, together with differential dust extinction curves and measurements of the 2175 {\AA} bump (already available for SBS 0909+532 \textit{AB} from \citealt{2002ApJ...574..719M}) will allow a detailed look at differences in dust structure and composition on kiloparsec scales. Increasing the samples of UV and optical high-resolution spectra for other gravitationally lensed quasar sight lines is also essential to understand how common the findings from SBS 0909+532 \textit{AB} are. Measurements of metallicities, relative abundances, dust depletions, and ionization parameters along the lensed sight lines for a large sample will improve understanding of the spatial scales of processes important for galaxy evolution, e.g., chemical enrichment and heating of the ISM by SNe Ia vs. SNe II supernovae and the processing of dust grains.

\acknowledgments
We thank an anonymous referee for constructive comments that have helped to improve this paper. F.H.C. and V.P.K. gratefully acknowledge support from NASA/ Space Telescope Science Institute (grant HST-GO-13801.001-A, PI Kulkarni) and from NASA grant NNX17AJ26G (PI Kulkarni). V.P.K. also gratefully acknowledges support from NASA grant 80NSSC20K0887 and NSF grant AST/2009811. S.L. was funded by project FONDECYT 1191232.

\vspace{5mm}
\facilities{HST(STIS), Keck(HIRES)}

\software{Cloudy 17.01 \citep{2017RMxAA..53..385F}, 
	  IRAF \citep{1993ASPC...52..173T},
          linetools \citep{2017zndo...1036773P},
          makee 6.4 (https://sites.astro.caltech.edu/~tb/makee/index.html),
          RDGEN 11.1 \citep{2014ascl.soft08017C},
          SPECP (D. Welty and J. Lauroesch),
          VPFIT 11.1 \citep{2014ascl.soft08015C}
          }


\clearpage
\begin{thebibliography}{}
\bibitem[Afruni et al.(2019)]{2019A&A...625A..11A} Afruni, A., Fraternali, F., \& Pezzulli, G.\ 2019, \aap, 625, A11
\bibitem[Andrews et al.(2001)]{2001ApJ...552L..73A} Andrews, S.~M., Meyer, D.~M., \& Lauroesch, J.~T.\ 2001, \apjl, 552, L73
\bibitem[Asplund et al.(2006)]{2006NuPhA.777....1A} Asplund, M., Grevesse, N., \& Jacques Sauval, A.\ 2006, \nphysa, 777, 1
\bibitem[Asplund et al.(2009)]{2009ARA&A..47..481A} Asplund, M., Grevesse, N., Sauval, A.~J., et al.\ 2009, \araa, 47, 481
\bibitem[Bergeron \& Stasi{\'n}ska(1986)]{1986A&A...169....1B} Bergeron, J., \& Stasi{\'n}ska, G.\ 1986, \aap, 169, 1
\bibitem[Bergeson et al.(1996)]{1996ApJ...464.1044B} Bergeson S.~D., Mullman K.~L., Wickliffe M.~E. et al.\ 1996 \apj, 464, 1044
\bibitem[Carswell et al.(2014)]{2014ascl.soft08017C} Carswell, R.~F., Webb, J.~K., Cooke, A.~J., et al.\ 2014, Astrophysics Source Code Library. ascl:1408.017
\bibitem[Carswell \& Webb(2014)]{2014ascl.soft08015C} Carswell, R.~F. \& Webb, J.~K.\ 2014, Astrophysics Source Code Library. ascl:1408.015
\bibitem[Cashman et al.(2017)]{2017ApJS..230....8C} Cashman F.~H., Kulkarni, V.~P., Kisielius, R. et al.\ 2017, \apjs, 230, 1
\bibitem[Cheng et al.(2012)]{2012ApJ...746..149C} Cheng, J.~Y., Rockosi, C.~M., Morrison, H.~L., et al.\ 2012, \apj, 746, 149
\bibitem[Churchill et al.(2003)]{2003ApJ...593..203C} Churchill, C.~W., Mellon, R.~R., Charlton, J.~C., et al.\ 2003, \apj, 593, 203. doi:10.1086/376444
\bibitem[Conroy et al.(2015)]{2015ApJ...803...77C} Conroy, C., van Dokkum, P.~G., \& Kravtsov, A.\ 2015, \apj, 803, 77
\bibitem[Cooke et al.(2010)]{2010MNRAS.409..679C} Cooke, R., Pettini, M., Steidel, C.~C., et al.\ 2010, \mnras, 409, 679. doi:10.1111/j.1365-2966.2010.17331.x
\bibitem[Dai \& Kochanek(2009)]{2009ApJ...692..677D} Dai, X., \& Kochanek, C.~S.\ 2009, \apj, 692, 677
\bibitem[Den Hartog et al.(2011)]{2011ApJS..194...35D} Den Hartog, E.~A., Lawler, J.~E., Sobeck, J.~S., et al.\ 2011, \apjs, 194, 35. doi:10.1088/0067-0049/194/2/35
\bibitem[Ferland et al.(2017)]{2017RMxAA..53..385F} Ferland, G.~J., Chatzikos, M., Guzm{\'a}n, F., et al.\ 2017, \rmxaa, 53, 385
\bibitem[Friel et al.(2002)]{2002AJ....124.2693F} Friel, E.~D., Janes, K.~A., Tavarez, M., et al.\ 2002, \aj, 124, 2693
\bibitem[Froese Fischer et al.(2006)]{2006ADNDT..92..607F} Froese Fischer C., Tachiev G. \& Irimia A.\ 2006, ADNDT, 92, 607
\bibitem[Fuhr \& Wiese(2006)]{2006JPCRD..35.1669F} Fuhr, J.~R. \& Wiese, W.~L.\ 2006, JPCRD, 35, 1669
\bibitem[Gauthier et al.(2010)]{2010ApJ...716.1263G} Gauthier, J.-R., Chen, H.-W., \& Tinker, J.~L.\ 2010, \apj, 716, 1263
\bibitem[Gauthier et al.(2009)]{2009ApJ...702...50G} Gauthier, J.-R., Chen, H.-W., \& Tinker, J.~L.\ 2009, \apj, 702, 50
\bibitem[Grossi et al.(2009)]{2009A&A...498..407G} Grossi, M., di Serego Alighieri, S., Giovanardi, C., et al.\ 2009, \aap, 498, 407
\bibitem[Hogg(1999)]{1999astro.ph..5116H} Hogg, D.~W.\ 1999, arXiv e-prints, astro-ph/9905116
\bibitem[Huang et al.(2016)]{2016MNRAS.455.1713H} Huang, Y.-H., Chen, H.-W., Johnson, S.~D., et al.\ 2016, \mnras, 455, 1713
\bibitem[Indriolo et al.(2007)]{2007ApJ...671.1736I} Indriolo, N., Geballe, T.~R., Oka, T., et al.\ 2007, \apj, 671, 1736. doi:10.1086/523036
\bibitem[Iwamoto et al.(1999)]{1999ApJS..125..439I} Iwamoto, K., Brachwitz, F., Nomoto, K., et al.\ 1999, \apjs, 125, 439
\bibitem[Keeton et al.(1998)]{1998ApJ...509..561K} Keeton, C.~R., Kochanek, C.~S., \& Falco, E.~E.\ 1998, \apj, 509, 561
\bibitem[Kennicutt et al.(2003)]{2003ApJ...591..801K} Kennicutt, R.~C., Bresolin, F., \& Garnett, D.~R.\ 2003, \apj, 591, 801
\bibitem[Khaire \& Srianand(2019)]{2019MNRAS.484.4174K} Khaire, V., \& Srianand, R.\ 2019, \mnras, 484, 4174
\bibitem[Kochanek et al.(1999)]{1999AIPC..470..163K} Kochanek, C.~S., Falco, E.~E., Impey, C.~D., et al.\ 1999, in AIP Conf. Proc.410, After the Dark Ages: When Galaxies Were Young (the Universe at 2 $< z <$ 5), ed. S. S. Holt (Melville, NY: AIP), 163 doi:10.1063/1.58598
\bibitem[Kochanek et al.(1997)]{1997ApJ...479..678K} Kochanek, C.~S., Falco, E.~E., Schild, R., et al.\ 1997, \apj, 479, 678
\bibitem[Koopmans et al.(2009)]{2009ApJ...703L..51K} Koopmans, L.~V.~E., Bolton, A., Treu, T., et al.\ 2009, \apjl, 703, L51
\bibitem[Krogager et al.(2018)]{2018A&A...619A.142K} Krogager, J.-K., Noterdaeme, P., O'Meara, J.~M., et al.\ 2018, \aap, 619, A142. doi:10.1051/0004-6361/201833608
\bibitem[Kulkarni et al.(2019)]{2019ApJ...886...83K} Kulkarni, V.~P., Cashman, F.~H., Lopez, S., et al.\ 2019, \apj, 886, 83
\bibitem[Lauroesch et al.(2000)]{2000ApJ...543L..43L} Lauroesch, J.~T., Meyer, D.~M., \& Blades, J.~C.\ 2000, \apjl, 543, L43
\bibitem[Leh{\'a}r et al.(2000)]{2000ApJ...536..584L} Leh{\'a}r, J., Falco, E.~E., Kochanek, C.~S., et al.\ 2000, \apj, 536, 584
\bibitem[Lopez et al.(2005)]{2005ApJ...626..767L} Lopez, S., Reimers, D., Gregg, M.~D., et al.\ 2005, \apj, 626, 767. doi:10.1086/429956
\bibitem[Loewenstein et al.(1998)]{1998ApJ...497..681L} Loewenstein, M., Hayashida, K., Toneri, T., et al.\ 1998, \apj, 497, 681
\bibitem[Lubin et al.(2000)]{2000AJ....119..451L} Lubin, L.~M., Fassnacht, C.~D., Readhead, A.~C.~S., et al.\ 2000, \aj, 119, 451
\bibitem[Luck \& Lambert(2011)]{2011AJ....142..136L} Luck, R.~E., \& Lambert, D.~L.\ 2011, \aj, 142, 136
\bibitem[Mathews \& Brighenti(2003)]{2003ARA&A..41..191M} Mathews, W.~G., \& Brighenti, F.\ 2003, \araa, 41, 191
\bibitem[Mediavilla et al.(2005)]{2005ApJ...619..749M} Mediavilla, E., Mu\~{n}oz, J.~A., Kochanek, C.~S. et al.\ 2005, \apj, 691, 749
\bibitem[Mernier et al.(2017)]{2017A&A...603A..80M} Mernier, F., de Plaa, J., Kaastra, J.~S., et al.\ 2017, \aap, 603, A80
\bibitem[Monier et al.(2009)]{2009MNRAS.397..943M} Monier, E.~M., Turnshek, D.~A., \& Rao, S.\ 2009, \mnras, 397, 943. doi:10.1111/j.1365-2966.2009.15000.x
\bibitem[Morton(2003)]{2003ApJS..149..205M} Morton D.~C.\ 2003, \apjs, 149, 205
\bibitem[Motta et al.(2002)]{2002ApJ...574..719M} Motta, V., Mediavilla, E., Mu{\~n}oz, J.~A., et al.\ 2002, \apj, 574, 719
\bibitem[Nelson et al.(2020)]{2020MNRAS.498.2391N} Nelson, D., Sharma, P., Pillepich, A., et al.\ 2020, \mnras, 498, 2391. doi:10.1093/mnras/staa2419
\bibitem[Oscoz et al.(1997)]{1997ApJ...491L...7O} Oscoz, A., Serra-Ricart, M., Mediavilla, E., et al.\ 1997, \apjl, 491, L7
\bibitem[Oosterloo et al.(2010)]{2010MNRAS.409..500O} Oosterloo, T., Morganti, R., Crocker, A., et al.\ 2010, \mnras, 409, 500
\bibitem[Pal\'{c}hikov(1998)]{1998PhyS...57..581P} Pal\'{c}hikov, V.~G.\ 1998, Phys. Scr., 57, 581 
\bibitem[Pipino \& Matteucci(2011)]{2011A&A...530A..98P} Pipino, A., \& Matteucci, F.\ 2011, \aap, 530, A98
\bibitem[Prochaska et al.(2003b)]{2018MNRAS.473.3559P} Prochaska J.~X., Gawiser E., Wolfe A.~M., et al.\ 2003b, \apj, 147, 227
\bibitem[Prochaska et al.(2017)]{2017zndo...1036773P} Prochaska, J.~X., Tejos, N., Crighton, N., et al.\ 2017, Zenodo, doi:10.5281/zenodo.1036773
\bibitem[Rao \& Turnshek(2000)]{2000ApJS..130....1R} Rao, S.~M. \& Turnshek, D.~A.\ 2000, \apj, 130, 1
\bibitem[Rosolowsky \& Simon(2008)]{2008ApJ...675.1213R} Rosolowsky, E., \& Simon, J.~D.\ 2008, \apj, 675, 1213
\bibitem[Rupke et al.(2010)]{2010ApJ...723.1255R} Rupke, D.~S.~N., Kewley, L.~J., \& Chien, L.-H.\ 2010, \apj, 723, 1255
\bibitem[Savage \& Sembach(1996)]{1996ARA&A..34..279S} Savage, B.~D. \& Sembach, K.~R. 1996, ARA\&A, 34, 279
\bibitem[Schneider et al.(1992)]{1992grle.book.....S} Schneider, P., Ehlers, J., \& Falco, E.~E.\ 1992, Gravitational Lenses (Berlin: Springer)
\bibitem[Serra et al.(2012)]{2012MNRAS.422.1835S} Serra, P., Oosterloo, T., Morganti, R., et al.\ 2012, \mnras, 422, 1835
\bibitem[Tody(1993)]{1993ASPC...52..173T} Tody, D.\ 1993, Astronomical Data Analysis Software and Systems II, 52, 173
\bibitem[Wucknitz et al.(2003)]{2003A&A...405..445W} Wucknitz, O., Wisotzki, L., Lopez, S., et al.\ 2003, \aap, 405, 445
\bibitem[Young et al.(2014)]{2014MNRAS.444.3408Y} Young, L.~M., Scott, N., Serra, P., et al.\ 2014, \mnras, 444, 3408
\bibitem[Zahedy et al.(2016)]{2016MNRAS.458.2423Z} Zahedy, F.~S., Chen, H.-W., Rauch, M., et al.\ 2016, \mnras, 458, 2423
\bibitem[Zahedy et al.(2017)]{2017ApJ...846L..29Z} Zahedy, F.~S., Chen, H.-W., Rauch, M., et al.\ 2017, \apjl, 846, L29

\end{thebibliography}
\end{document}